\documentclass[reprint,floatfix]{revtex4-1}
\usepackage{amsmath}
\usepackage{booktabs} 
\usepackage{bm}
\usepackage{makecell}
\usepackage{graphicx}
\usepackage{xcolor}
\draft 

\begin{document}


\title{Regularized Fluctuating Lattice Boltzmann Model} 



\author{M. Lauricella}
\email{marco.lauricella@cnr.it}
\thanks{Corresponding author}
\affiliation{Istituto per le Applicazioni del Calcolo CNR, via dei Taurini 19, 00185 Rome, Italy}
\author{A. Montessori}
\affiliation{Department of Civil, Computer Science and Aeronautical Technologies Engineering, Roma Tre University, Via Vito Volterra, Rome, 00146, Italy}

\author{A. Tiribocchi}
\affiliation{Istituto per le Applicazioni del Calcolo CNR, via dei Taurini 19, 00185 Rome, Italy}
\affiliation{INFN "Tor Vergata" Via della Ricerca Scientifica 1, 00133 Rome, Italy}
\author{S. Succi}
\affiliation{Center for Life Nano Science@La Sapienza, Istituto Italiano di Tecnologia, 00161 Roma, Italy}
\affiliation{Department of Physics, Harvard University, Cambridge, MA, 02138, USA}
\affiliation{Istituto per le Applicazioni del Calcolo CNR, via dei Taurini 19, 00185 Rome, Italy}


\date{\today}

\begin{abstract}
We introduce a regularized fluctuating lattice Boltzmann model (Reg-FLBM) for the D3Q27 lattice, which incorporates thermal fluctuations through Hermite-based projections to ensure compliance with the fluctuation–dissipation theorem. By leveraging the recursive regularization framework, the model achieves thermodynamic consistency for both hydrodynamic and ghost modes. Compared to the conventional single-relaxation-time BGK-FLBM, the Reg-FLBM provides improved stability and a more accurate description of thermal fluctuations. The implementation is optimized for large-scale parallel simulations on GPU-accelerated architectures, enabling systematic investigation of fluctuation-driven phenomena in mesoscale and nanoscale fluid systems.
\end{abstract}

\pacs{}

\maketitle 

\section{Introduction}

Over the past few decades, the lattice Boltzmann method (LBM) has emerged as a powerful alternative for solving the deterministic Navier–Stokes equations from lattice kinetic theory \cite{tiribocchi2025lattice,succi2018lattice,kruger2017lattice,aidun2010lattice,benzi1992lattice}. A clear prove is the large employment of LBM to model complex hydrodynamic phenomena, such as fluids exhibiting phase transitions and/or phase separation due to non-ideal interactions \cite{montessori2025thread,zhang2024improved,tiribocchi2020novel,chiappini2019hydrodynamic,montessori2019mesoscale,liu2012three,he1999lattice}, particle suspensions \cite{pelusi2023sharp,guglietta2023suspensions,bonaccorso2020lbsoft,nguyen2002lubrication,ladd2001lattice}, polymer flows \cite{monteferrante2021lattice,malaspinas2010lattice,berk2005lattice,ahlrichs1999simulation,ahlrichs1998lattice}, electrohydrodynamic problems \cite{xiong2025thermodynamically,liu2024consistent,wang2021lattice,lauricella2018entropic,kupershtokh2006lattice}, reactive flows \cite{sawant2021lattice,lin2017multi}, active matter \cite{tiribocchi2023crucial,carenza2020chaotic,de2016lattice,marenduzzo2007steady}, just
to name notable examples. The success of LBM is largely due to its high degree of physical flexibility and computational efficiency across the
full spectrum of scales of motion. Indeed, alongside its capability of capturing the physics of fluid interfaces with great accuracy, LBM leverages extreme scalability on parallel architectures, a feature that,
together with the rapid development of GPU-based high-performance computing (HPC) clusters, has enabled the simulation of increasingly complex systems at high spatial and temporal resolutions on multi-GPUs platforms \cite{lauricella2025acclb,latt2025multi,tiribocchi2023lightweight,patel2023imexlbm}. 

Thanks to this improved computational capacity, LBM simulations can now resolve fluid dynamics down to nanometric scales \cite{zhao2023minireview}, enabling the investigation of phenomena where continuum assumptions begin to break down and molecular-level effects, such as thermal fluctuations, can play a fundamental role in determining the macroscopic behavior of the system. While thermal fluctuations are typically negligible in classical continuum models at larger spatial resolutions, this assumption no longer holds at nanometric scales. In such regimes, fluctuations become an intrinsic component of the fluid dynamics and must be explicitly incorporated into the model to ensure physical consistency \cite{forster2018hydrodynamic,zwanzig2001nonequilibrium,landau1987fluid}.

The incorporation of thermal fluctuations is generally more straightforward in Lagrangian models, where stochastic forces can be directly added to particle trajectories in accordance with the fluctuation-dissipation theorem \cite{frenkel2023understanding}.  
In this context notable examples are the Velocity rescaling
algorithm
\cite{bussi2008stochastic,bussi2007canonical}, the Langevin dynamics \cite{tanygin2024comparison,bussi2007accurate} and the Nosé–Hoover chains \cite{martyna1992nose}, to name a few.

Within the LBM framework, incorporating fluctuations into the mesoscopic description typically requires the introduction of stochastic fluxes or stresses, which must be handled carefully to preserve the thermodynamic consistency of the fluid. A major step forward in modeling fluctuating hydrodynamics with LBM was first achieved in the simulation of particle suspensions in fluids in Refs. \cite{ladd1993short,ladd1994numerical}, where 
hydrodynamic variables were linked to the statistical fluctuations of the lattice Boltzmann populations, laying the foundation for the so-called fluctuating lattice Boltzmann method (FLBM) \cite{dunweg2007statistical,adhikari2005fluctuating}. In particular, Adhikari et al. \cite{adhikari2005fluctuating} demonstrated that stochasticity must be extended to all non-conserved modes, each corresponding to a moment obtained by projecting the population vector onto an element of an orthonormal basis, to maintain consistency with the fluctuation–dissipation theorem (FDT), especially at finite wave numbers.

The FLBM has evolved into a robust simulation framework capable of handling both ideal and non-ideal, single- and multicomponent fluids, exploiting both free-energy-based formulations \cite{gross2011modelling} and the pseudopotential approach \cite{belardinelli2015fluctuating,sbragaglia2013interaction}. These models reveal the emergence of scaling laws in density correlations and nonequilibrium-induced pressure corrections \cite{belardinelli2019lattice}.
Such methodological advances have extended the applicability of  FLBM to fluctuating diffusion equations \cite{wagner2016fluctuating}, while algorithmic improvements have been introduced to reinforce Galilean invariance \cite{kaehler2013fluctuating}. The effects of geometric confinement were also investigated by Xue et al. \cite{xue2021lattice}, showing that the presence of walls can significantly amplify velocity fluctuations in multicomponent fluids up to an order of magnitude compared to unconfined systems.

In the present work, we introduce a regularized fluctuating lattice Boltzmann model (Reg-FLBM), where thermal fluctuations are included following the fluctuation amplitudes derived by Dünweg et al. \cite{dunweg2007statistical}, an approach ensuring consistency with the fluctuation–dissipation theorem. However, unlike traditional implementations, the orthonormal basis used to define the moments is constructed following the Hermite polynomial basis proposed by O. Malaspinas in the context of the D3Q27 lattice \cite{malaspinas2015increasing}. \textcolor{black}{The D3Q27 lattice belongs to the class of full Hermite schemes, as it provides an exact Gauss–Hermite quadrature up to second order, thereby ensuring a complete representation of all Hermite polynomials required for the recovery of isothermal hydrodynamics \cite{coreixas2017recursive}.} This choice enables the model to retain the advantages of regularized lattice Boltzmann approaches, including recursive formulations for the equilibrium and non--equilibrium distribution functions, as well as improved stability properties \cite{mattila2017high}. By leveraging this Hermite basis, fluctuations can be consistently projected onto the non-hydrodynamic subspace while maintaining compatibility with regularization LB procedures introduced in the literature \cite{coreixas2017recursive,latt2006lattice}.

This paper is structured as follows: In Section \ref{sec:method}, we illustrate the idea of the regularized FLBM, while in Section III, we present some test cases to validate the method. Finally, in Section IV, we present performance data, while in Section V, we outline conclusions and perspectives.

\section{METHOD}
\label{sec:method}

\subsection{Fluctuating BGK lattice Boltzmann}
\label{subsec:flbgk}

In this subsection, we outline the main ingredients of the lattice Boltzmann framework, with particular focus on the fluctuating BGK formulation.

The second-order accurate Bhatnagar–Gross–Krook (BGK) lattice Boltzmann equation (LBE) defined over a discrete velocity set \( \mathbf{c}_i\) reads
\begin{align}
f_i(x_\alpha + c_{i\alpha}\Delta t, t + \Delta t) &= f_i(x_\alpha,t) - \omega \left(f_i - f^{\text{eq}}_i\right) \notag \\
&= f^{\text{eq}}_i + (1 - \omega) f^{\text{neq}}_i, \label{lbe}
\end{align}
where \( \omega = 1/\tau \) is the relaxation frequency related to the kinematic viscosity by \( \nu = c_s^2(\tau - 0.5) \) with \( c_s^2 = 1/3 \) in lattice units, \( f_i \) denote the distribution functions (or populations) defined along the discrete velocities \( \mathbf{c}_i \), while \( f^{\text{eq}}_i \) are their equilibrium counterparts given as Mach expansion of the Maxwell--Boltzmann distribution of the molecular velocity~\cite{succi2018lattice,kruger2017lattice},
whose explicit form is:
\begin{equation}\label{classicfeq} 
f_i^{eq}
= w_i\,\rho\left[
1+\frac{c_{i\alpha}u_\alpha}{c_s^2}
+\frac{(c_{i\alpha}u_\alpha)^2}{2c_s^4}
-\frac{u^2}{2c_s^2}
\right],
\end{equation} 
where the Einstein summation convention over repeated indices is assumed ($\alpha,\beta = x,y,z$).

In the second equality of Eq. \ref{lbe}, we exploit $f_i=f^{eq}_i+f^{neq}_i$.
Finally, the Greek indices denote Cartesian components. 
Hydrodynamic quantities (such as density and momentum) are computed as zeroth and first moments of the populations
\begin{align}
\rho &= \sum_i f_i(\mathbf{x}, t), \label{eq:rho} \\
\rho u_\alpha &= \sum_i f_i(\mathbf{x}, t) c_{i\alpha}, \label{eq:u}
\end{align}
where ${\bf u}$ is the fluid velocity. 

Following Refs.\cite{adhikari2005fluctuating,dunweg2007statistical,schiller2008thermal}, thermal fluctuations can be included via a noise term, satisfying the fluctuation-dissipation
theorem (FDT) at the lattice level,
on the right-hand side of Eq.\ref{lbe}. To do this, it is convenient to follow the multi-relaxation time model (MRT)\cite{d2002multiple,lallemand2000theory}, where the distribution functions \( f_i \) are projected  onto a complete set of moments (or modes) \( m_k \), defined as
\begin{equation} \label{moments}
m_k = \sum_{i} e_{ki} f_i.
\end{equation}
Here, \( \mathbf{e}_k = (e_{k0}, e_{k1}, \dots, e_{k26}) \) are orthogonal vectors (detailed in the next subsection) in the D3Q27 representation satisfying \( \sum_i w_i e_{ki} e_{li} = b_k \delta_{kl} \), \( w_i \) are weights (see Table~\ref{tab:d3q27}) and \( b_k \) are normalization factors, defined as
\begin{equation} \label{normal_fact}
b_k = \sum_i w_i e_{ki}^2.
\end{equation}

\begin{table}[h]
\centering
\caption{Discrete velocities and weights for the D3Q27 lattice.}
\label{tab:d3q27}
\begin{tabular}{c c c c}
\hline
\textbf{i} & \textbf{Velocity} $(c_{ix}, c_{iy}, c_{iz})$ & $\|\mathbf{c}_i\|^2$ & $w_i$ \\
\hline
0 & $(0,0,0)$ & 0 & $8/27$ \\
1--6 & $(\pm1,0,0), (0,\pm1,0), (0,0,\pm1)$ & 1 & $2/27$ \\
7--18 & $(\pm1,\pm1,0), (\pm1,0,\pm1), (0,\pm1,\pm1)$ & 2 & $1/54$ \\
19--26 & $(\pm1,\pm1,\pm1)$ & 3 & $1/216$ \\
\hline
\end{tabular}
\end{table}

The basis vectors \( \mathbf{e}_k \) are designed to correspond to physically meaningful degrees of freedom such as mass, momentum, stress, and higher-order modes~\cite{d2002multiple}. 
Following the procedure adopted for $f_i$,
each moment \( m_k \) can be decomposed into equilibrium and non--equilibrium parts
\[
m_k = m_k^{\text{eq}} + m_k^{\text{neq}},
\]
where both terms are obtained by applying Eq.~\ref{moments} to \( f_i^{\text{eq}} \) and \( f_i^{\text{neq}} \), respectively.
Note that Eq.\ref{lbe}  can be decomposed into a collision step followed by a streaming step  as follows
\begin{align}
  & f_i^*(x_\alpha,t) = f_i^{\text{eq}}(x_\alpha,t) + (1 - \omega) f_i^{\text{neq}}(x_\alpha,t), & \label{eq:collision} \\
 & f_i(x_\alpha + c_{i\alpha} \Delta t, t + \Delta t) = f_i^*(x_\alpha,t), & \label{eq:streaming}
\end{align}
where the superscript \( ^* \) denotes post-collision values.

Applying Eq.~\ref{moments} to the post-collision non--equilibrium distribution \( f_i^{\text{neq}*} \) yields the post-collision moments~\cite{schiller2008thermal}
\begin{equation} 
m_k^{\text{neq},*} = (1 - \omega) \, m_k^{\text{neq}} + \varphi_k r_k,
\end{equation}
where \( \varphi_k r_k \) is a stochastic term added to the non-conserved modes (\( k = 4,\dots,26 \)) only. Here, \( r_k \) is a standard normal random variable and \( \varphi_k \) is the fluctuation amplitude ensuring thermodynamic consistency via the fluctuation–dissipation theorem.

According to Ref.~\cite{dunweg2007statistical}, the stochastic term in the moment collision operator can be interpreted as a Monte Carlo process satisfying detailed balance, hence the noise amplitude reads
\begin{equation} \label{fluct_phi}
\varphi_k = \sqrt{ \frac{\rho\, k_B T\, \omega (2 - \omega)\, b_k}{c_s^2} },
\end{equation}
where \( k_B T \) sets the fluctuation variance of the non-conserved modes.

The distribution functions can then be obtained via a back-transformation\cite{dunweg2007statistical,schiller2008thermal}
\begin{equation} \label{backtrasf}
f_i = w_i \sum_k b_k^{-1} e_{ki} m_k,
\end{equation}
which is the inverse of Eq.~\ref{moments}, based on the orthogonality of \( \mathbf{e}_k \).
Finally, substituting the noisy moments into the BGK update
yields the fluctuating lattice Boltzmann equation
\begin{equation}\label{flbe}
\begin{split}
f_i(x_\alpha + c_{i\alpha} \Delta t, t + \Delta t)
  &= f_i^{\text{eq}} + (1 - \omega) f_i^{\text{neq}} \\
  &\quad + w_i \sum_{k=4}^{26} b_k^{-1} e_{ki} \varphi_k r_k .
\end{split}
\end{equation}

This formulation defines the FLBM with additive thermal noise applied to the non-conserved modes, ensuring consistency with fluctuating hydrodynamics in the low Mach number limit.

\subsection{Regularized fluctuating lattice Boltzmann}
\label{subsec:Reg-fluct}

In this subsection, we summarize the essential aspects of the regularized lattice Boltzmann method (Reg-LBM), with particular emphasis on the Hermite-based formulation of the equilibrium distribution function. \textcolor{black}{Briefly, in the Reg-LBM, the 
non--equilibrium contribution is not taken simply as 
$f^{\text{neq}} = f - f^{\text{eq}}$, but is reconstructed 
by projecting the distribution onto a Hermite basis \cite{coreixas2017recursive,malaspinas2015increasing}. 
This procedure separates the hydrodynamic modes from the 
higher-order ghost modes in a consistent way.}

To this aim, following the approach outlined in foundational works on the regularization procedure~\cite{latt2006lattice,zhang2006efficient,ladd2001lattice}, the distribution functions \( f_i \) can be formally expanded around their equilibrium value
\begin{equation}
\label{eq:expanded}
f_i = f_i^{(0)} + f_i^{(1)} + f_i^{(2)} + \dots,
\end{equation}
where \( f_i^{(0)} \equiv f_i^{eq} \) is the equilibrium distribution function, and the higher-order terms \( f_i^{(k)} \) represent corrections of order \( O(\epsilon^k) \), with \( \epsilon \ll 1 \) denoting the Knudsen number.

By applying a multiscale Chapman–Enskog expansion~\cite{chapman1990mathematical}, it is shown that retaining only the first two terms, \( f_i^{(0)} \) and \( f_i^{(1)} \), is sufficient to asymptotically recover the Navier–Stokes equations~\cite{kruger2017lattice,latt2008straight}. Accordingly
\begin{equation}
f_i = f_i^{(0)} + f_i^{(1)} + O(\epsilon^2),
\end{equation}
and the non--equilibrium contribution can be approximated as
\begin{equation}
f_i^{(1)} = f_i - f_i^{(0)} + O(\epsilon^2) \approx f_i^{neq}.
\end{equation}
Thus, to leading order, the distribution reads
\begin{equation}
f_i = f_i^{eq} + f_i^{neq}.
\end{equation}

Both \( f_i^{(0)} \) and \( f_i^{(1)} \) can be expanded over Hermite polynomials up to an arbitrary order \( n \) ~\cite{coreixas2017recursive,malaspinas2015increasing,shan2006kinetic,grad1949note,grad1949kinetic}
\begin{equation} \label{hermite_feq}
f_i^{(0)} = w_i \sum_n \frac{1}{c_s^{2n} \, n!} \, a^{(n)}_{0, \alpha_1 \ldots \alpha_n} \, \mathcal{H}^{(n)}_{i \alpha_1 \ldots \alpha_n},
\end{equation}
\begin{equation} \label{hermite_fneq}
f_i^{(1)} = w_i \sum_n \frac{1}{c_s^{2n} \, n!} \, a^{(n)}_{1, \alpha_1 \ldots \alpha_n} \, \mathcal{H}^{(n)}_{i \alpha_1 \ldots \alpha_n}.
\end{equation}

Here, \( \mathcal{H}^{(n)} \) denotes the \( n \)-th order Hermite polynomial tensor, and \( a^{(n)}_{0,\ldots} \), \( a^{(n)}_{1,\ldots} \) are the Hermite coefficients of the equilibrium and non--equilibrium parts, respectively. The indices \( \alpha_1 \ldots \alpha_n \) run over the spatial directions \( x, y, z \), indicating that the Hermite coefficients and basis functions are rank-\( n \) tensors defined over the 3D velocity space. \textcolor{black}{Note that, when truncated at second order in the Hermite expansion, $f_i^{(0)}$ reduces exactly to the standard equilibrium distribution $f_i^{eq}$ reported in Eq. \ref{classicfeq}.}

\begin{table*}[ht]
\caption{Hermite basis functions for the D3Q27 stencil. For each $k$-th vector, we report the Hermite symbol $H^{(n)}_{\dots}$, the corresponding polynomial expression $e_{ki}$, the polynomial order $n_k$, its multiplicity $\mu_k$, and the normalization factor $b_k$. The components of the discrete velocities $\mathbf{c}_i$ are denoted as $c_{i\alpha}$ with $\alpha = x, y, z$.}
\centering
\resizebox{0.85\textwidth}{!}{%
\begin{tabular}{cccccc}
\hline
$k$ &  $H^{(n)}_{\dots}$ & $e_{ki}$ & $n_k$ & $\mu_k$ & $b_k$ \\
\hline
0  & $H^{(0)}$   & $1$ & 0 & 1 & $1$ \\
1  & $H^{(1)}_x$   & $c_{ix}$ & 1 & 1 & $ 1/3  $ \\
2  & $H^{(1)}_y$   & $c_{iy}$ & 1 & 1 & $ 1/3  $ \\
3  & $H^{(1)}_z$   & $c_{iz}$ & 1 & 1 & $ 1/3  $ \\
4  & $H^{(2)}_{xx}$  & $c_{ix}^2 - c_s^2$ & 2 & 1 & $ 2/9  $ \\
5  & $H^{(2)}_{yy}$  & $c_{iy}^2 - c_s^2$ & 2 & 1 & $ 2/9  $ \\
6  & $H^{(2)}_{zz}$  & $c_{iz}^2 - c_s^2$ & 2 & 1 & $ 2/9  $ \\
7  & $H^{(2)}_{xy}$  & $c_{ix} c_{iy}$ & 2 & 2 & $ 1/9  $ \\
8  & $H^{(2)}_{xz}$  & $c_{ix} c_{iz}$ & 2 & 2 & $ 1/9  $ \\
9  & $H^{(2)}_{yz}$  & $c_{iy} c_{iz}$ & 2 & 2 & $ 1/9  $ \\
10 & $H^{(3)}_{xxy}$ & $c_{ix}^2 c_{iy} - c_s^2 c_{iy}$ & 3 & 3 & $ 2/27  $ \\
11 & $H^{(3)}_{xxz}$ & $c_{ix}^2 c_{iz} - c_s^2 c_{iz}$ & 3 & 3 & $ 2/27  $ \\
12 & $H^{(3)}_{xyy}$ & $c_{ix} c_{iy}^2 - c_s^2 c_{ix}$ & 3 & 3 & $ 2/27  $ \\
13 & $H^{(3)}_{xzz}$ & $c_{ix} c_{iz}^2 - c_s^2 c_{ix}$ & 3 & 3 & $ 2/27  $ \\
14 & $H^{(3)}_{yzz}$ & $c_{iy} c_{iz}^2 - c_s^2 c_{iy}$ & 3 & 3 & $ 2/27  $ \\
15 & $H^{(3)}_{yyz}$ & $c_{iy}^2 c_{iz} - c_s^2 c_{iz}$ & 3 & 3 & $ 2/27  $ \\
16 & $H^{(3)}_{xyz}$ & $c_{ix} c_{iy} c_{iz}$ & 3 & 6 & $ 1/27  $ \\
17 & $H^{(4)}_{xxyy}$ & $c_{ix}^2 c_{iy}^2 - c_s^2 (c_{ix}^2 + c_{iy}^2) + c_s^4$ & 4 & 6 & $ 4/81  $ \\
18 & $H^{(4)}_{xxzz}$ & $c_{ix}^2 c_{iz}^2 - c_s^2 (c_{ix}^2 + c_{iz}^2) + c_s^4$ & 4 & 6 & $ 4/81  $ \\
19 & $H^{(4)}_{yyzz}$ & $c_{iy}^2 c_{iz}^2 - c_s^2 (c_{iy}^2 + c_{iz}^2) + c_s^4$ & 4 & 6 & $ 4/81  $ \\
20 & $H^{(4)}_{xyzz}$ & $c_{ix} c_{iy} c_{iz}^2 - c_s^2 c_{ix} c_{iy}$ & 4 & 12 & $ 2/81  $ \\
21 & $H^{(4)}_{xyyz}$ & $c_{ix} c_{iy}^2 c_{iz} - c_s^2 c_{ix} c_{iz}$ & 4 & 12 & $ 2/81  $ \\
22 & $H^{(4)}_{xxyz}$ & $c_{ix}^2 c_{iy} c_{iz} - c_s^2 c_{iy} c_{iz}$ & 4 & 12 & $ 2/81  $ \\
23 & $H^{(5)}_{x^2yz^2}$ & $c_{ix}^2 c_{iy} c_{iz}^2 - c_s^2 (c_{ix}^2 c_{iy} + c_{iy} c_{iz}^2) + c_s^4 c_{iy}$ & 5 & 30 & $ 4/243  $ \\
24 & $H^{(5)}_{x^2y^2z}$ & $c_{ix}^2 c_{iy}^2 c_{iz} - c_s^2 (c_{ix}^2 c_{iz} + c_{iy}^2 c_{iz}) + c_s^4 c_{iz}$ & 5 & 30 & $ 4/243  $ \\
25 & $H^{(5)}_{xy^2z^2}$ & $c_{ix} c_{iy}^2 c_{iz}^2 - c_s^2 (c_{ix} c_{iy}^2 + c_{ix} c_{iz}^2) + c_s^4 c_{ix}$ & 5 & 30 & $ 4/243  $ \\
26 & $H^{(6)}_{x^2y^2z^2}$ & $c_{ix}^2 c_{iy}^2 c_{iz}^2 - c_s^2 (c_{ix}^2 c_{iy}^2 + c_{ix}^2 c_{iz}^2 + c_{iy}^2 c_{iz}^2) + c_s^4 (c_{ix}^2 + c_{iy}^2 + c_{iz}^2) - c_s^6$ & 6 & 90 & $ 8/729  $ \\
\hline
\end{tabular}
}
\label{tab:hermite_d3q27_cs}
\end{table*}

In this work, we adopt as basis vector $e_{ki}$ the orthogonal Hermite basis specifically constructed for the D3Q27 lattice~\cite{malaspinas2015increasing}, consisting of 27 polynomial functions forming a complete and orthogonal set under the discrete scalar product weighted by the lattice weights \( w_i \), with normalization factors \( b_k \) as defined in Table~\ref{tab:hermite_d3q27_cs}. The orthogonality allows coefficients to be computed via projections
\begin{equation} \label{eq:hermitcoeffeq}
a^{(n)}_{0,\alpha_1 \ldots \alpha_n} = \sum_i f^{(0)}_i \, \mathcal{H}^{(n)}_{i, \alpha_1 \ldots \alpha_n},
\end{equation}
\begin{equation} \label{eq:hermitcoeffneq}
a^{(n)}_{1,\alpha_1 \ldots \alpha_n} = \sum_i f^{(1)}_i \, \mathcal{H}^{(n)}_{i, \alpha_1 \ldots \alpha_n}, \quad \text{for } n \geq 2.
\end{equation}
It is worth noting that \( a_1^{(n)} \) is defined only for \( n \geq 2 \), since mass and momentum (associated with \( n = 0,1 \)) are conserved and do not admit non--equilibrium corrections.

A key advantage of the orthogonal Hermite basis lies in its recursive construction of high-order coefficients
\begin{equation}\label{rec-eq}
a^{(n)}_{0,\alpha_1 \ldots \alpha_n} = a^{(n-1)}_{0,\alpha_1 \ldots \alpha_{n-1}} \, u_{\alpha_n}, \quad a^{(0)}_0 = \rho,
\end{equation}
\begin{align}\label{recursive_eq}
a^{(n)}_{1,\alpha_1 \ldots \alpha_n}
  &= a^{(n-1)}_{1,\alpha_1 \ldots \alpha_{n-1}} \, u_{\alpha_n} \nonumber\\
  &\quad + \left( u_{\alpha_1} \ldots u_{\alpha_{n-2}}
      a^{(2)}_{1,\alpha_{n-1} \alpha_n}
      + \text{perm}(\alpha_n) \right).
\end{align}
where “perm(\( \alpha_n \))” denotes cyclic permutations over the first \( n-1 \) indices. This construction reduces memory requirements and simplifies the implementation of the regularization procedure.

The normalization factors are connected to the orthogonal Hermite basis as
\begin{equation}
b_k = \frac{n_k! \cdot c_s^{2n_k}}{\mu_k},
\end{equation}
where \( n_k \) is the order of the \( k \)-th basis function, and \( \mu_k \) its multiplicity, as reported in Table~\ref{tab:hermite_d3q27_cs}.

Hence, the discrete LBE with regularization (Reg-LBM) can be augmented
with the noise terms, becoming the Reg-FLBM:
\begin{equation}\label{r-flbe}
\begin{split}
f_i(x_\alpha + c_{i\alpha} \Delta t, t + \Delta t)
  &= f_i^{(0)} + (1 - \omega) f_i^{(1)} \\
  &\quad + w_i \sum_{k=4}^{26} b_k^{-1} e_{ki} \varphi_k r_k .
\end{split}
\end{equation}

To decouple hydrodynamic and ghost contributions, we split $f_i^{(1)}$ as
\[
f_i^{(1)} = f_i^{(1,h)} + f_i^{(1,g)},
\]
and introduce distinct relaxation rates \( \omega_h \) and \( \omega_g \) for the non-conserved hydrodynamics modes \( k=4\dots9 \), and the ghost modes \( k=10\dots26 \), respectively. Following Refs~\cite{dunweg2007statistical,adhikari2005fluctuating}, we set \( \omega_g = 1 \) to avoid spurious dissipation of thermal noise by ghost modes.

Thus, the final form of the regularized fluctuating LBE reads
\begin{equation}\label{r-flbe2}
\begin{split}
f_i(x_\alpha + c_{i\alpha} \Delta t, t + \Delta t)
  &= f_i^{(0)} + (1 - \omega_h) f_i^{(1,h)} \\
  &\quad + w_i \sum_{k=4}^{26} b_k^{-1} e_{ki} \varphi_k r_k ,
\end{split}
\end{equation}
where \( \varphi_k \) incorporates the relaxation rate \( \omega_k \) of the \( k \)-th mode as defined in Eq.~\ref{fluct_phi}. Also,
\begin{equation} \label{hermite_feq2}
    f_i^{(0)} = w_i \sum_{k=0}^{26} \frac{\mu_k}{c_s^{2n_k} \, n_k!} \, a^{(n_k)}_{0, \alpha_1 \ldots \alpha_{n_k}} \, \mathcal{H}^{(n_k)}_{i \alpha_1 \ldots \alpha_{n_k}},
\end{equation}
\begin{equation} \label{hermite_fneq2h}
f_i^{(1,h)} = w_i \sum_{k=4}^{9} \frac{\mu_k}{c_s^{2n_k} \, n_k!} \, a^{(n_k)}_{1, \alpha_1 \ldots \alpha_{n_k}} \, \mathcal{H}^{(n_k)}_{i \alpha_1 \ldots \alpha_{n_k}},
\end{equation}
with the equilibrium part \( f_i^{(0)} \) evaluated using the recursive relation in Eq.~\ref{recursive_eq}.
Interestingly, Eqs.~(\ref{hermite_feq2}) and (\ref{hermite_fneq2h}) provide a component-wise expression of the back-transformation defined in Eq.~\ref{backtrasf}, confirming the consistency of the Hermite-based expansion
\textcolor{black}{, where the transformation matrix from population
space to moment space reads  $M_{ki}=e_{ki}$ and the inverse matrix is $M^{-1}_{ki}=w_i b_k^{-1} e_{ki}$.}

\textcolor{black}{To summarize, by employing a full Hermite representation on the D3Q27 lattice, and using the recursive relations in Eqs.~\ref{recursive_eq}, we are able to construct both hydrodynamic and non-hydrodynamic (ghost) Hermite moments. 
In this way, higher-order ghost modes are expressed explicitly as combinations of lower-order hydrodynamic moments, thereby integrating fluctuations consistently within the same Hermite framework.}

\textcolor{black}{Finally, for completeness, we recall the microscopic picture underlying this formulation.  
At the kinetic level, the single–particle distribution in velocity space is the Maxwell--Boltzmann, and the LB equilibrium, $f_i^{(0)} \equiv f_i^{\rm eq}$, is its Gauss--Hermite projection on the discrete set of velocities given in Eq. \ref{hermite_feq2}. Instead, the \emph{Poisson} ingredient pertains to \emph{number fluctuations} (shot noise) in a particle picture, 
as typically encountered in integer lattice gas models \cite{seekins2025integer,blommel2018integer}: stochastic derivations and integer lattice gases lead to a counting process whose covariances are proportional to the mean occupancy, yielding in the ideal gas limit the discrete relation $\langle \delta f_i\,\delta f_j\rangle = f^{\rm eq}_i\,\delta_{ij}$ when expressed on the LB velocity set. In practical FLBM implementations, one does not sample Poisson occupancies; rather, one adds \emph{Gaussian} noise with amplitudes fixed by the FDT so as to reproduce the equilibrium variances of hydrodynamic fields. This Gaussian forcing corresponds to the large–occupancy limit of the Poisson shot noise: deviations from strict Poisson statistics may arise only at very low effective particle numbers per cell and vanish rapidly upon increasing the
coarse–graining.}

\begin{table*}[ht]
\caption{Space-and time-averaged quadratic fluctuations of the Equilibration Ratio (ER) for selected hydrodynamic quantities, as a function of the relaxation time $\tau_h$. The main values are computed using the regularized fluctuating lattice Boltzmann model (Reg-FLBM). In contrast, values in parentheses correspond to results obtained with the BGK fluctuating model (BGK-FLBM) where available.}
\centering
\begin{tabular}{c|cccc}
\toprule
\hline
$\tau$ & $\mathrm{ER}(\rho)$ & $\mathrm{ER}(\sum_\alpha \rho u_{\alpha})$ & $\mathrm{ER}(\sum_{\alpha\alpha} \Pi_{\alpha\alpha})$ & $\mathrm{ER}(\sum_{\alpha<\beta} \Pi_{\alpha\beta})$ \\
\hline
\midrule
0.5001 & 1.037 (\dots) & 1.052 (\dots) & 1.072 (\dots) & 1.048 (\dots) \\
0.5005 & 1.013 (\dots) & 1.032 (\dots) & 1.035 (\dots) & 1.031 (\dots) \\
0.501  & 1.010 (\dots) & 1.026 (\dots) & 1.022 (\dots) & 1.025 (\dots) \\
0.505  & 1.006 (1.027) & 1.015 (1.058) & 1.005 (1.034) & 1.016 (1.082) \\
0.51   & 1.005 (1.014) & 1.013 (1.020) & 1.004 (1.005) & 1.012 (1.030) \\
0.55   & 1.003 (1.006) & 1.007 (1.008) & 1.005 (1.007) & 1.005 (1.007) \\
0.7    & 1.002 (1.003) & 1.003 (1.004) & 1.004 (1.004) & 1.003 (1.004) \\
1.0    & 1.002 (1.002) & 1.002 (1.002) & 1.003 (1.003) & 1.003 (1.003) \\
1.5    & 1.002 (1.001) & 1.002 (1.001) & 1.003 (1.003) & 1.003 (1.002) \\
2.0   & 1.002 (1.001) & 1.002 (1.001) & 1.003 (1.002) & 1.003 (1.002) \\
5.0    & 1.001 (1.000) & 1.001 (1.000) & 1.003 (1.001) & 1.003 (1.001) \\
10.0   & 1.001 (0.991) & 1.001 (1.001) & 1.002 (0.987) & 1.003 (1.001) \\
50.0   & 1.000 (0.953) & 1.001 (1.001) & 1.002 (0.930) & 1.003 (1.000) \\
100.0  & 1.000 (0.953) & 1.001 (1.002) & 1.002 (0.929) & 1.003 (1.001) \\
\bottomrule
\hline
\end{tabular}
\label{tab:flucts}
\end{table*}

\section{RESULTS}
\label{sec:results}

To validate the implementation of the fluctuating lattice Boltzmann models in the multi-GPU based \textit{accLB} code\cite{lauricella2025acclb}, we performed a systematic set of simulations on a cubic domain of size $256^3$ lattice nodes. We choose mass, length, and time units such that the density $\rho$ is normalized to $1$ on a unit lattice. Each simulation was evolved for 500,000 time steps, while density $\rho$, momentum $\rho \bm{u}$, and the momentum flux tensor components $\Pi_{\alpha\beta}$ were saved every 5,000 steps, yielding 100 temporal snapshots per run.
We tested both the classic BGK-based fluctuating LBM (FLBM-BGK), as formulated in Eq. \ref{flbe}, and the regularized fluctuating model (Reg-FLBM), as expressed in Eq. \ref{r-flbe2}. 
\textcolor{black}{In order to ensure a consistent comparison, in both cases the equilibrium distribution was defined through Eq.~\ref{hermite_feq2}, so that the only difference between the two approaches lies in the treatment of the non--equilibrium contribution: in the BGK case it is simply obtained as $f^{\text{neq}}_i = f_i - f^{\text{eq}}_i$, while in the Reg-FLBM it is reconstructed by projecting onto the full Hermite basis.}
This setup provides a clean benchmark to assess the impact of regularization on numerical stability and thermodynamic consistency across a broad range of relaxation parameters.

\begin{figure*}[ht]
  \centering
\includegraphics[width=0.95\textwidth]{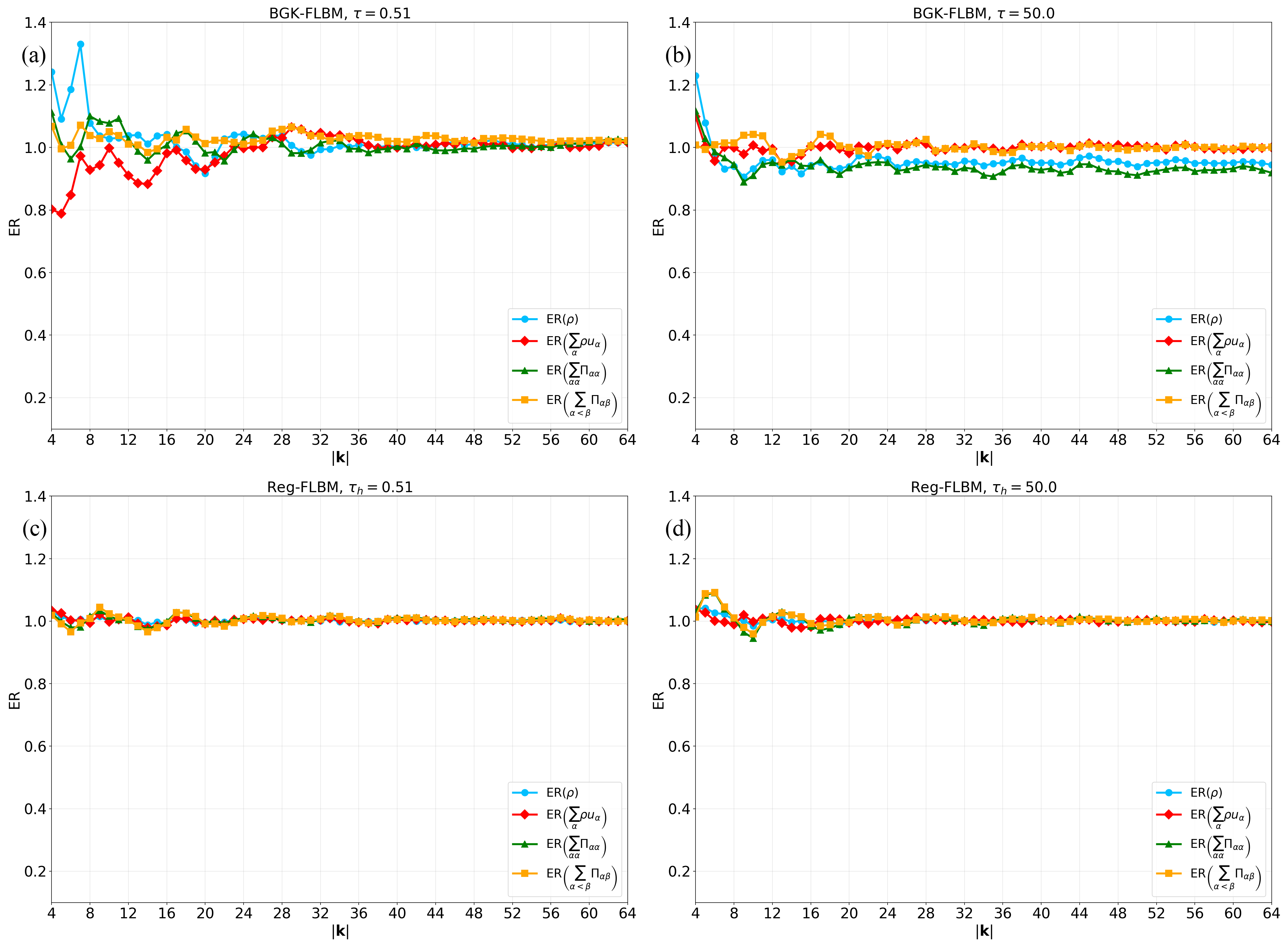}
  \caption{Equilibration ratio spectra for the four main hydrodynamic fields, computed as a function of the wavevector modulus $k$, for both the BGK-FLBM (top row) and the Reg-FLBM (bottom row), at two values of the relaxation time: $\tau=0.51$ (left column) and $\tau=50.0$ (right column). Panels: (a) BGK-FLBM, $\tau=0.51$; (b) BGK-FLBM, $\tau=50.0$; (c) Reg-FLBM, $\tau=0.51$; (d) Reg-FLBM, $\tau=50.0$.}
  \label{fig:fft_comparison}
\end{figure*}

For each model, simulations were carried out using a fixed thermal energy of $k_BT = 1/3000$, in line with previous studies~\cite{adhikari2005fluctuating,dunweg2007statistical}. We explored a broad range of relaxation times $\tau$. Specifically, for the Reg-FLBM, we investigated values of the hydrodynamics relaxation time $\tau_h=1/\omega_h$ spanning from 0.5001 (vanishing viscosity regime) to 100, thereby probing both under-relaxation and over-relaxation regimes. 
In contrast, for the FLBM-BGK model, the smallest numerically stable value for the single relaxation time was $\tau = 0.505$, consistent with known limitations in the literature regarding numerical stability of BGK-based fluctuating solvers \cite{ollila2011fluctuating,bernaschi2009muphy}. The use of high-viscosity regimes ($\tau \gg  1$) in our study is motivated by several factors. From a physical perspective, this limit is particularly relevant for rarefied gases, where the dynamics are characterized by long mean free paths and the crossover to the ballistic regime becomes significant. The regularized lattice Boltzmann method offers enhanced numerical stability in the under-relaxation regime by using a recursive high-order Hermite expansion and systematically reprojecting the non--equilibrium part onto the correct Hermite modes, which would otherwise lead to unphysical artifacts and instability in standard BGK-based schemes \cite{feng2019hybrid,montessori2018regularized}.

First, for fluctuating hydrodynamic quantities, we numerically assess the ``equilibration ratio'' (ER), defined as the ratio between the numerically measured variance of a given hydrodynamic field and the corresponding theoretical value predicted by equilibrium statistical mechanics \cite{landau1987fluid}. It provides a quantitative measure of how accurately thermal fluctuations are reproduced in the simulation \cite{adhikari2005fluctuating}. 

First, for fluctuating hydrodynamic quantities, we numerically assess the
\emph{equilibration ratio} (ER), defined as the ratio between the numerically
measured variance of a given hydrodynamic field and the corresponding theoretical
value predicted by equilibrium statistical mechanics~\cite{landau1987fluid}.
It provides a quantitative measure of how accurately thermal fluctuations are
reproduced in the simulation~\cite{adhikari2005fluctuating}.


\textcolor{black}{The Equilibration Ratio (ER) is a fluctuating observable, which we define as
\begin{equation}
\mathrm{ER}(m) = 
\frac{\langle (\delta m)^2 \rangle}{\langle (\delta m)^2 \rangle_{\text{theory}}},
\end{equation}
where $m$ denotes a given fluctuating mode with $\delta m = (m - \langle m \rangle)$  (e.g., density, momentum, or stress tensor component), 
$\langle (\delta m)^2 \rangle$ is the variance obtained from the simulation, 
and $\langle (\delta m)^2 \rangle_{\text{theory}}$ is the theoretical prediction from the fluctuation--dissipation theorem \cite{landau1987fluid}.  Thus, $ER=1$ indicates perfect agreement with equilibrium statistical mechanics, while deviations quantify incomplete thermalization.}

In Table~\ref{tab:flucts} we report the space-and time-averaged fluctuations for the ``equilibration ratio'' for density, $\mathrm{ER}(\rho)$, momentum, $\mathrm{ER}(\sum_\alpha \rho u_{\alpha})$, diagonal components of the momentum flux tensor, $\mathrm{ER}(\sum_{\alpha\alpha} \Pi_{\alpha\alpha} )$,
and off-diagonal components, $\mathrm{ER}(\sum_{\alpha<\beta} \Pi_{\alpha\beta})$, where the terms of the momentum flux tensor are computed as $\Pi_{\alpha\beta}=a^{(2)}_{0,\alpha\beta}+a^{(2)}_{1,\alpha\beta}$ by Eq.  \ref{eq:hermitcoeffeq} and Eq. \ref{eq:hermitcoeffneq}.
The results clearly demonstrate the high accuracy and robustness of the Reg-FLBM scheme across a wide range of relaxation times $\tau$. In particular, Reg-FLBM yields quadratic fluctuations within $0.5\%$ of the theoretical prediction for all measured hydrodynamic quantities in the range $0.51 \lesssim \tau \lesssim 100$, and for all fields outside the over-relaxation (vanishing viscosity) regime. This excellent agreement is maintained even as $\tau$ is increased up to 100, where all ER remain extremely close to unity. Thus, the regularization ensures excellent agreement with theoretical predictions even for extreme values of $\tau \gg 1$, enabling stable and accurate simulations across the entire spectrum of viscosities. In contrast, the standard FLBM-BGK results (shown in parentheses) exhibit significantly larger discrepancies, especially for the diagonal components of the momentum flux tensor at high viscosities (i.e., in the under-relaxing regime), where deviations from the theoretical value can exceed $7\%$. 

At low viscosities (over-relaxing regime, $\tau \to 0.5$), both models encounter the well-established limits set by numerical stability. The FLBM-BGK model is only marginally stable down to $\tau=0.505$, where the ER for the diagonal stress already shows discrepancies as large as $5\%$ and the statistics become noisy for all fields. In contrast, the Reg-FLBM maintains satisfactory reproduction of equilibrium fluctuations at these values, with deviations close to $1\%$ for all fields, and remains numerically stable even down to $\tau=0.5001$. For the lowest accessible viscosity ($\tau=0.5001$), deviations in the diagonal pressure term approach $7\%$, but the density and momentum remain well equilibrated. These findings highlight the improved numerical stability and thermodynamic consistency of the regularized scheme, particularly in the vanishing viscosity (over-relaxation) limit, and its clear superiority over the classical BGK-FLBM for accurately reproducing hydrodynamic fluctuations across all regimes.

To assess whether the amplitude of thermal fluctuations is correctly reproduced across all spatial scales, we performed a spatial Fourier transform of the hydrodynamic fields at each snapshot, \textcolor{black}{obtained from a normalized FFT}. For each wavevector $\mathbf{k}$, the fluctuation spectra were then averaged over all wavevectors with the same modulus $|\mathbf{k}|$, i.e., a spherical average in Fourier space, in order to obtain the isotropic spectrum $S(|\mathbf{k}|)= \langle (\delta m(|\mathbf{k}|))^2 \rangle$. This approach enables us to resolve the fluctuation spectra as a function of the wavevector magnitude $|\mathbf{k}|$, thereby directly verifying the agreement between the numerically measured and theoretically expected variances of each mode. By additionally averaging the spherical spectra over all snapshots in time, we obtain a robust estimate of the stationary fluctuation spectrum. Such an analysis, inspired by Adhikari et al.~\cite{adhikari2005fluctuating}, enables us to check the accuracy of the fluctuation–dissipation balance not just for spatially averaged quantities, but along a broad range of correlation lengths, corresponding to different $k$-modes in the system. In particular, we focused on the analysis of the spectral amplitudes at selected wavevector magnitudes $|\mathbf{k}| = 4$ and $|\mathbf{k}| = 64$, avoiding very small or very large $|\mathbf{k}|$ where discretization or finite-size effects may become dominant.

Figure~\ref{fig:fft_comparison} shows the $|\mathbf{k}|$-dependent equilibration ratios for four hydrodynamic fields, density ($\mathrm{ER}(\rho)$), total momentum ($\mathrm{ER}(\sum_\alpha \rho u_\alpha)$), diagonal momentum flux tensor ($\mathrm{ER}(\sum_{\alpha} \Pi_{\alpha\alpha})$), and off-diagonal momentum flux tensor ($\mathrm{ER}(\sum_{\alpha<\beta} \Pi_{\alpha\beta})$) as a function of the wavevector modulus $|\mathbf{k}|$. Results are presented for both the BGK-FLBM and the Reg-FLBM at two representative values of the relaxation parameter: $\tau = 0.51$ (low viscosity/over-relaxation) and $\tau = 50.0$ (high viscosity/under-relaxation).

Comparing the panels reveals several trends. For the BGK-FLBM (top row), at low viscosity, $\tau = 0.51$ in panel (a), the density and diagonal stress components exhibit noticeable deviations from unity at small $|\mathbf{k}|$, with significant scatter and systematic underestimation in the diagonal pressure sector (triangle symbol), in agreement with the limitations discussed in Adhikari et al.\cite{adhikari2005fluctuating} and Dünweg et al.\cite{dunweg2007statistical}. At high viscosity, $\tau=50.0$ in panel (b), these discrepancies become more pronounced and persistent across a broad range of $|\mathbf{k}|$, particularly for the stress tensor, reflecting the difficulty of the BGK scheme in thermalizing all non-hydrodynamic modes.

For the Reg-FLBM, the bottom row in panels (c) and (d) shows a much more robust agreement with theoretical predictions across all $|\mathbf{k}|$ values, both at low and high viscosity values. All four fields display equilibration ratios that remain close to unity, with only minor residual fluctuations at very small $|\mathbf{k}|$. This demonstrates the improved thermodynamic consistency of the regularized scheme, which effectively ensures the correct amplitude of equilibrium fluctuations for all hydrodynamic and ghost modes even in the extreme under- and over-relaxation regimes.

Overall, these results confirm that the Reg-FLBM consistently reproduces the theoretical equilibrium fluctuation spectra for all hydrodynamic quantities, whereas the BGK-FLBM exhibits significant departures from the expected behavior, especially for the stress tensor and at small $k$ in the over-relaxation regime.

\begin{figure}[ht]
  \centering
 \includegraphics[width=0.95\columnwidth]{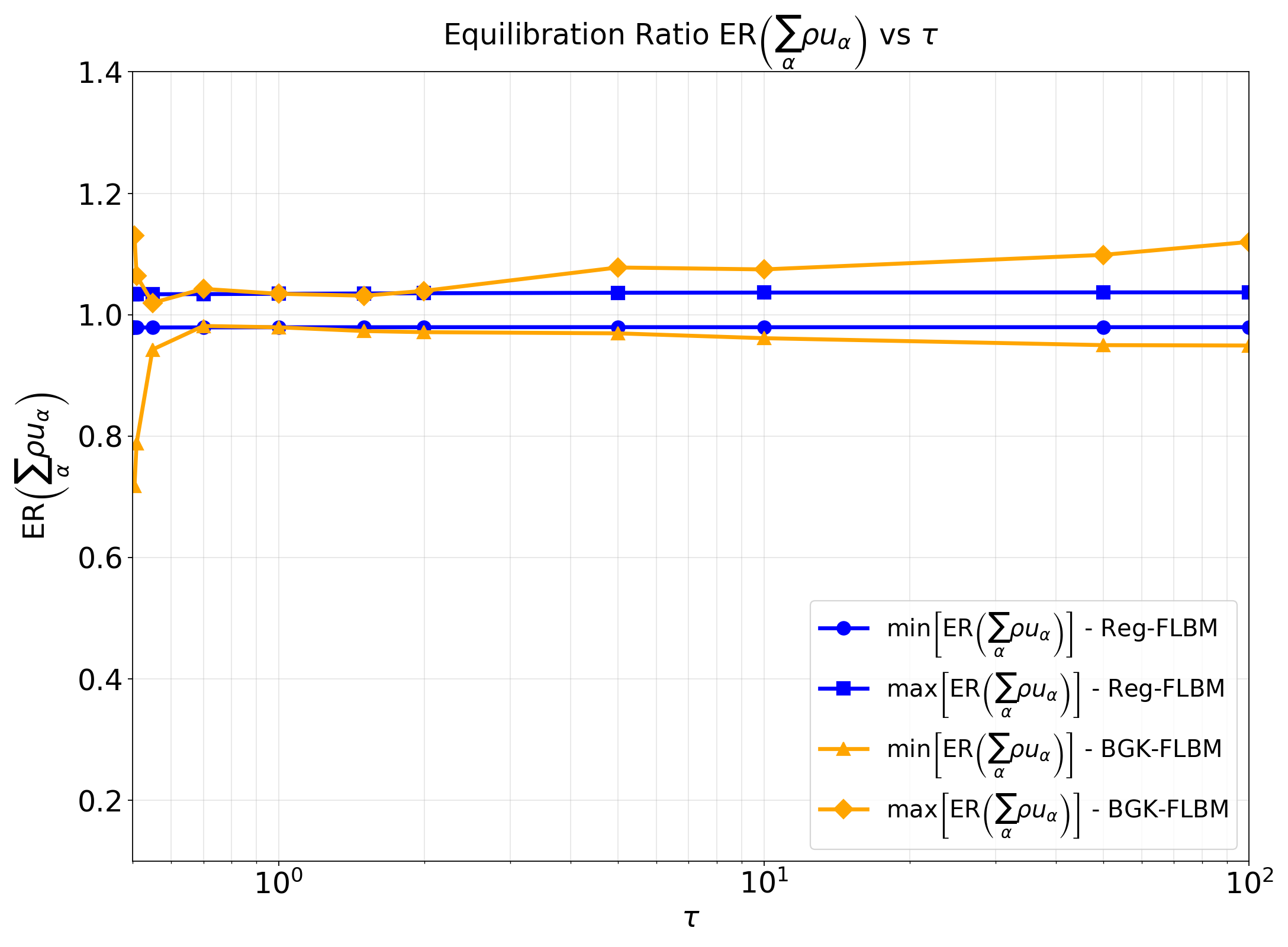}
  \caption{
    Minimum and maximum values of the equilibration ratio for the total momentum, $\mathrm{ER}\left(\sum_\alpha \rho u_\alpha\right)$, as a function of the relaxation time $\tau$, for the Reg-FLBM (blue, circles and squares) and the BGK-FLBM (orange, triangles and diamonds). For each value of $\tau$, the minimum and maximum deviation from unity are computed over all wavevector magnitudes $|\mathbf{k}|$ in the range $4 \leq |\mathbf{k}| \leq 64$.
  }
  \label{fig:minmax_comparison}
\end{figure}

To highlight the discrepancies between the two models across the explored relaxation times, Figure~\ref{fig:minmax_comparison} shows the minimum and maximum values of the equilibration ratio for the total momentum, $\mathrm{ER}(\sum_\alpha \rho u_\alpha)$, as a function of the relaxation time $\tau$, for both the Reg-FLBM and the BGK-FLBM. The Reg-FLBM results (blue curves) display a remarkable stability and remain close to unity for all values of $\tau$, with very small spread between the minimum and maximum across the entire range. In contrast, the BGK-FLBM (orange curves) exhibits significantly larger deviations, especially at low $\tau$, where the spread between minimum and maximum values becomes pronounced, indicating a lack of proper thermalization of the momentum fluctuations in this regime. These results highlight the superior thermodynamic consistency of the regularized scheme over the standard BGK implementation within the framework of FLBM.

\section{PERFORMANCE}

To evaluate the parallel performance of the code, we performed weak scaling benchmarks in which each GPU handles a fixed sub-domain of $256^3$ lattice sites, and the global simulation box is extended along the $z$ direction as $L_z = 256 \times n_\mathrm{GPU}$. All simulations were run with the multi-gpu based \textit{accLB} code\cite{lauricella2025acclb} on the Leonardo supercomputer at CINECA, using NVIDIA Ampere™ A100 GPUs and hybrid MPI+OpenACC parallelization.

We compare the regularized fluctuating Lattice Boltzmann model (Reg-FLBM) with its deterministic counterpart (Reg-LBM), where the regularization is retained but thermal fluctuations are disabled. Performance is measured in GLUPS (Giga Lattice Updates Per Second), defined as
\[
\mathrm{GLUPS} = \frac{L_x L_y L_z}{10^9 \, t_s},
\]
where $t_s$ is the average wall-clock time per simulation step.

\begin{figure}[ht]
  \centering
 \includegraphics[width=0.95\columnwidth]{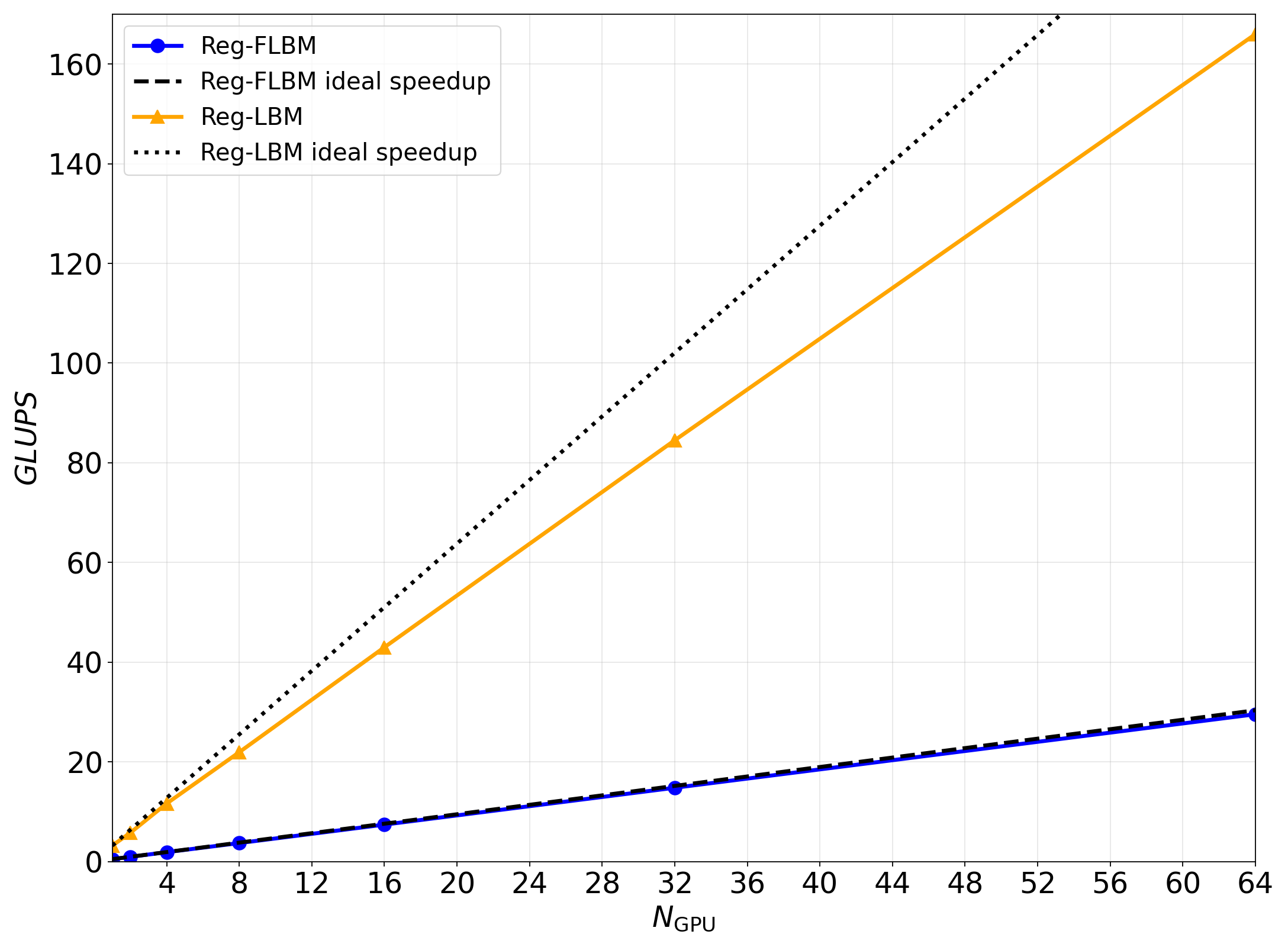}
  \caption{Weak scaling of GLUPS (Giga Lattice Updates Per Second) as a function of the number of GPUs, $N_{\mathrm{GPU}}$, for the regularized fluctuating (Reg-FLBM) and regularized non-fluctuating (Reg-LBM) models on Leonardo at CINECA. Each GPU handles a $256^3$ subdomain. Dashed and dotted lines show ideal speedup. Both models achieve nearly ideal scaling up to 64 GPUs, with Reg-FLBM reflecting the additional cost of fluctuations.}
  \label{fig:glups}
\end{figure}

Figure~\ref{fig:glups} presents the weak scaling behavior of both the Reg-FLBM and the deterministic Reg-LBM on the Leonardo supercomputer. For each case, the workload per GPU is kept constant at $256^3$ lattice sites by proportionally increasing the global $z$-extent with the number of GPUs, following a linear decomposition. Both models exhibit nearly ideal scaling up to 64 GPUs, with the measured GLUPS growing linearly with $N_{\mathrm{GPU}}$ and closely tracking the ideal speedup line.

The performance gap between the fluctuating and deterministic models remains approximately constant as the system is scaled, reflecting the extra computational cost of generating and applying random numbers at each site for the fluctuating case. This overhead results in a sustained per-GPU performance reduction of approximately a factor of 6–7, but does not affect the parallel efficiency, which remains high for both models.

The observed scaling is enabled by the thread-safe implementation of the algorithm, which prevents memory race conditions and allows for fully local, independent updates across all threads~\cite{montessori2023thread}. By reconstructing the post-collision distributions through Hermite projection, the code achieves efficient coalesced memory access and optimal data locality. The choice of a structure-of-arrays (SoA) memory layout further improves bandwidth utilization, supporting high throughput and scalability across modern GPU-based architectures.

These results demonstrate that the code architecture is well-suited for massively parallel simulations, achieving both high absolute performance and excellent scalability, which are essential for large-scale, fluctuation-resolving lattice Boltzmann studies in high-performance computing environments.

\section{CONCLUSIONS AND PERSPECTIVES}

In this work, we have introduced and validated a regularized fluctuating lattice Boltzmann model (Reg-FLBM) based on a Hermite polynomial basis for the D3Q27 lattice. Our approach provides a consistent and robust framework for incorporating thermal fluctuations into lattice Boltzmann simulations, ensuring compatibility with the fluctuation–dissipation theorem and leveraging the advantages of recursive regularization schemes. Extensive numerical tests confirm that the Reg-FLBM accurately reproduces the equilibrium variance of hydrodynamic fields across a wide range of relaxation times, showing deviations of less than 0.5\% from the theoretical predictions for all measured quantities, except in the extreme over-relaxation regime.

Compared to the classic BGK-FLBM, the regularized model exhibits markedly improved thermodynamic consistency and numerical stability, particularly for the diagonal stress components and at very low viscosities, where the BGK approach is known to suffer from unphysical noise amplification and loss of equilibrium. Spectral analyses further demonstrate that the Reg-FLBM correctly reproduces the fluctuation amplitude at all spatial scales, confirming the correct implementation of the fluctuation–dissipation balance for both hydrodynamic and ghost modes.

On the computational side, large-scale benchmarks on the Leonardo supercomputer demonstrate that the Reg-FLBM exhibits excellent weak scaling up to 64 GPUs, achieving high parallel efficiency due to its thread-safe design and memory layout. Although the inclusion of thermal noise incurs a computational overhead of about a factor of 6–7 compared to the non-fluctuating Reg-LBM, the code architecture fully exploits the capabilities of modern GPU clusters, making it suitable for fluctuation-resolving simulations of complex fluids at unprecedented scale and fidelity.

In perspective, moving to high-order lattices (such as the full Hermite scheme D3Q125) is expected to further enhance numerical stability and isotropy. This improvement stems from the fact that the local curvature induced by a given force $\mathbf{F}$ over a time step, scaling as $\mathbf{F}/(m\,\mathbf{c})\cdot\Delta t$, becomes smaller as the lattice velocity discretization $\mathbf{c}$ increases. As a result, for the same force, the population deflection per step is reduced, leading to smoother dynamics and mitigating the onset of spurious anisotropies and ghost-mode artifacts, which become especially critical in the under-relaxed regime ($\tau \gg 1$). Moreover, higher-order lattices provide a more isotropic representation of hydrodynamic stresses, further enhancing simulation accuracy in highly viscous or force-driven flows\cite{malaspinas2015increasing,coreixas2017recursive}.

Further, the present framework can be naturally extended to multicomponent and non-ideal fluid systems, where thermal fluctuations play a key role in phase separation, interface dynamics, and critical phenomena. Previous studies have shown that fluctuating lattice Boltzmann approaches can be generalized to include non-ideal interactions and multiple species, ensuring consistent thermalization of both hydrodynamic and ghost modes \cite{belardinelli2015fluctuating,sbragaglia2013interaction,gross2011modelling}. 

\textcolor{black}{In this context, it has been pointed out that the fluctuation spectrum of non-ideal fluids may involve spatially correlated noise rather than the uncorrelated Poissonian form valid for the ideal-gas case \cite{gross2011modelling}. More recently, Parsa and Wagner~\cite{parsa2020large} demonstrated how coarse-graining can give rise to correlated noise contributions in non-ideal systems, leading to a richer fluctuation structure. Addressing these effects within the present regularization framework remains an open problem and represents a natural direction for future developments.}
These extensions pave the way for systematic investigations of fluctuation-driven phenomena in binary mixtures and non-ideal fluids, with broad applications spanning from soft condensed matter to micro- and nanofluidics.

In addition, the framework presented here could be combined with a fluctuating lattice Boltzmann approach for the diffusion or temperature equation, following the double-population scheme flows in the framework of the Boussinesq assumption~\cite{d2003boundary,guo2002coupled}. Recent advances in fluctuating LBM for diffusive equations~\cite{wagner2016fluctuating} open the way to fully consistent simulations of fluctuating hydrodynamics and thermodynamics, where both velocity and temperature (or concentration) fields include thermal noise in accordance with the fluctuation–dissipation theorem.
\textcolor{black}{Furthermore, in more complex fluids such as binary mixtures, liquid crystals, or active matter, fluctuations also act on additional order parameters, which may be scalar, vectorial, or tensorial in nature. In such cases, it is natural to consider hybrid strategies that combine fluctuating LB for the fluid dynamics with finite-difference or finite-volume discretizations of the accompanying fluctuating PDEs. These couplings can be realized efficiently, although the balance between accuracy and computational cost will depend on the particular system and the level of coarse-graining, and, therefore, offer a relevant avenue for future studies.}

Finally, future improvements in the Galilean invariance of fluctuating lattice Boltzmann schemes may be achieved by adopting velocity-dependent orthogonalization and noise projection procedures, as proposed by Kaehler and Wagner \cite{kaehler2013fluctuating}. Incorporating such advanced treatments of noise ensures a consistent fluctuation–dissipation balance and minimizes spurious violations of Galilean invariance, especially in flows with non-zero mean velocities. \textcolor{black}{ In addition, a recent work by Feng and co-workers \cite{feng2019hybrid} has demonstrated that Hermite-based regularization itself reduces Galilean invariance violations, thereby further supporting the present approach as a promising framework 
for fluctuation-resolving simulations in moving systems.}

In light of these perspectives, the Reg-FLBM provides a robust, accurate, and scalable tool for studying hydrodynamic fluctuations in mesoscale and nanoscale fluid systems, laying the groundwork for systematic investigations of fluctuation-dominated phenomena.

\section*{Data Availability}
The data that support the findings of this study are available from the corresponding author upon reasonable request.

\section*{Dedication}
This work is dedicated to our colleague and friend Michele Parrinello on the occasion of his 80th birthday, with our warmest wishes to keep going with game-changing ideas for many years to come.

\begin{acknowledgments}
M.L. and A.M. acknowledge funding from the Italian Government through the PRIN (Progetti di Rilevante Interesse Nazionale) Grant (MOBIOS) ID: 2022N4ZNH3 -CUP: F53C24001000006 and computational support of CINECA through the ISCRA B project DODECA (HP10BUBFIL). M.L. and A. T. acknowledge the support of the Italian National Group for Mathematical Physics (GNFM-INdAM). M.L., A.T., and S.S. acknowledge the support from the European Research Council under the ERCPoC Grant No. 101187935 (LBFAST).
\end{acknowledgments}


\begin{thebibliography}{81}%
\makeatletter
\providecommand \@ifxundefined [1]{%
 \@ifx{#1\undefined}
}%
\providecommand \@ifnum [1]{%
 \ifnum #1\expandafter \@firstoftwo
 \else \expandafter \@secondoftwo
 \fi
}%
\providecommand \@ifx [1]{%
 \ifx #1\expandafter \@firstoftwo
 \else \expandafter \@secondoftwo
 \fi
}%
\providecommand \natexlab [1]{#1}%
\providecommand \enquote  [1]{``#1''}%
\providecommand \bibnamefont  [1]{#1}%
\providecommand \bibfnamefont [1]{#1}%
\providecommand \citenamefont [1]{#1}%
\providecommand \href@noop [0]{\@secondoftwo}%
\providecommand \href [0]{\begingroup \@sanitize@url \@href}%
\providecommand \@href[1]{\@@startlink{#1}\@@href}%
\providecommand \@@href[1]{\endgroup#1\@@endlink}%
\providecommand \@sanitize@url [0]{\catcode `\\12\catcode `\$12\catcode
  `\&12\catcode `\#12\catcode `\^12\catcode `\_12\catcode `\%12\relax}%
\providecommand \@@startlink[1]{}%
\providecommand \@@endlink[0]{}%
\providecommand \url  [0]{\begingroup\@sanitize@url \@url }%
\providecommand \@url [1]{\endgroup\@href {#1}{\urlprefix }}%
\providecommand \urlprefix  [0]{URL }%
\providecommand \Eprint [0]{\href }%
\providecommand \doibase [0]{http://dx.doi.org/}%
\providecommand \selectlanguage [0]{\@gobble}%
\providecommand \bibinfo  [0]{\@secondoftwo}%
\providecommand \bibfield  [0]{\@secondoftwo}%
\providecommand \translation [1]{[#1]}%
\providecommand \BibitemOpen [0]{}%
\providecommand \bibitemStop [0]{}%
\providecommand \bibitemNoStop [0]{.\EOS\space}%
\providecommand \EOS [0]{\spacefactor3000\relax}%
\providecommand \BibitemShut  [1]{\csname bibitem#1\endcsname}%
\let\auto@bib@innerbib\@empty
\bibitem [{\citenamefont {Tiribocchi}\ \emph {et~al.}(2025)\citenamefont
  {Tiribocchi}, \citenamefont {Durve}, \citenamefont {Lauricella},
  \citenamefont {Montessori}, \citenamefont {Tucny},\ and\ \citenamefont
  {Succi}}]{tiribocchi2025lattice}%
  \BibitemOpen
  \bibfield  {author} {\bibinfo {author} {\bibfnamefont {A.}~\bibnamefont
  {Tiribocchi}}, \bibinfo {author} {\bibfnamefont {M.}~\bibnamefont {Durve}},
  \bibinfo {author} {\bibfnamefont {M.}~\bibnamefont {Lauricella}}, \bibinfo
  {author} {\bibfnamefont {A.}~\bibnamefont {Montessori}}, \bibinfo {author}
  {\bibfnamefont {J.-M.}\ \bibnamefont {Tucny}}, \ and\ \bibinfo {author}
  {\bibfnamefont {S.}~\bibnamefont {Succi}},\ }\href@noop {} {\bibfield
  {journal} {\bibinfo  {journal} {Physics Reports}\ }\textbf {\bibinfo {volume}
  {1105}},\ \bibinfo {pages} {1} (\bibinfo {year} {2025})}\BibitemShut
  {NoStop}%
\bibitem [{\citenamefont {Succi}(2018)}]{succi2018lattice}%
  \BibitemOpen
  \bibfield  {author} {\bibinfo {author} {\bibfnamefont {S.}~\bibnamefont
  {Succi}},\ }\href@noop {} {\emph {\bibinfo {title} {The lattice Boltzmann
  equation: for complex states of flowing matter}}}\ (\bibinfo  {publisher}
  {Oxford university press},\ \bibinfo {year} {2018})\BibitemShut {NoStop}%
\bibitem [{\citenamefont {Kr{\"u}ger}\ \emph {et~al.}(2017)\citenamefont
  {Kr{\"u}ger}, \citenamefont {Kusumaatmaja}, \citenamefont {Kuzmin},
  \citenamefont {Shardt}, \citenamefont {Silva},\ and\ \citenamefont
  {Viggen}}]{kruger2017lattice}%
  \BibitemOpen
  \bibfield  {author} {\bibinfo {author} {\bibfnamefont {T.}~\bibnamefont
  {Kr{\"u}ger}}, \bibinfo {author} {\bibfnamefont {H.}~\bibnamefont
  {Kusumaatmaja}}, \bibinfo {author} {\bibfnamefont {A.}~\bibnamefont
  {Kuzmin}}, \bibinfo {author} {\bibfnamefont {O.}~\bibnamefont {Shardt}},
  \bibinfo {author} {\bibfnamefont {G.}~\bibnamefont {Silva}}, \ and\ \bibinfo
  {author} {\bibfnamefont {E.~M.}\ \bibnamefont {Viggen}},\ }\href@noop {}
  {\bibfield  {journal} {\bibinfo  {journal} {Springer International
  Publishing}\ }\textbf {\bibinfo {volume} {10}},\ \bibinfo {pages} {4}
  (\bibinfo {year} {2017})}\BibitemShut {NoStop}%
\bibitem [{\citenamefont {Aidun}\ and\ \citenamefont
  {Clausen}(2010)}]{aidun2010lattice}%
  \BibitemOpen
  \bibfield  {author} {\bibinfo {author} {\bibfnamefont {C.~K.}\ \bibnamefont
  {Aidun}}\ and\ \bibinfo {author} {\bibfnamefont {J.~R.}\ \bibnamefont
  {Clausen}},\ }\href@noop {} {\bibfield  {journal} {\bibinfo  {journal}
  {Annual review of fluid mechanics}\ }\textbf {\bibinfo {volume} {42}},\
  \bibinfo {pages} {439} (\bibinfo {year} {2010})}\BibitemShut {NoStop}%
\bibitem [{\citenamefont {Benzi}, \citenamefont {Succi},\ and\ \citenamefont
  {Vergassola}(1992)}]{benzi1992lattice}%
  \BibitemOpen
  \bibfield  {author} {\bibinfo {author} {\bibfnamefont {R.}~\bibnamefont
  {Benzi}}, \bibinfo {author} {\bibfnamefont {S.}~\bibnamefont {Succi}}, \ and\
  \bibinfo {author} {\bibfnamefont {M.}~\bibnamefont {Vergassola}},\
  }\href@noop {} {\bibfield  {journal} {\bibinfo  {journal} {Physics Reports}\
  }\textbf {\bibinfo {volume} {222}},\ \bibinfo {pages} {145} (\bibinfo {year}
  {1992})}\BibitemShut {NoStop}%
\bibitem [{\citenamefont {Montessori}, \citenamefont {Hegele},\ and\
  \citenamefont {Lauricella}(2025)}]{montessori2025thread}%
  \BibitemOpen
  \bibfield  {author} {\bibinfo {author} {\bibfnamefont {A.}~\bibnamefont
  {Montessori}}, \bibinfo {author} {\bibfnamefont {L.~A.}\ \bibnamefont
  {Hegele}}, \ and\ \bibinfo {author} {\bibfnamefont {M.}~\bibnamefont
  {Lauricella}},\ }\href@noop {} {\bibfield  {journal} {\bibinfo  {journal}
  {AIAA Journal}\ }\textbf {\bibinfo {volume} {63}},\ \bibinfo {pages} {1005}
  (\bibinfo {year} {2025})}\BibitemShut {NoStop}%
\bibitem [{\citenamefont {Zhang}\ \emph {et~al.}(2024)\citenamefont {Zhang},
  \citenamefont {Li}, \citenamefont {Wang},\ and\ \citenamefont
  {Shu}}]{zhang2024improved}%
  \BibitemOpen
  \bibfield  {author} {\bibinfo {author} {\bibfnamefont {D.}~\bibnamefont
  {Zhang}}, \bibinfo {author} {\bibfnamefont {Y.}~\bibnamefont {Li}}, \bibinfo
  {author} {\bibfnamefont {Y.}~\bibnamefont {Wang}}, \ and\ \bibinfo {author}
  {\bibfnamefont {C.}~\bibnamefont {Shu}},\ }\href@noop {} {\bibfield
  {journal} {\bibinfo  {journal} {Physics of Fluids}\ }\textbf {\bibinfo
  {volume} {36}} (\bibinfo {year} {2024})}\BibitemShut {NoStop}%
\bibitem [{\citenamefont {Tiribocchi}\ \emph {et~al.}(2020)\citenamefont
  {Tiribocchi}, \citenamefont {Montessori}, \citenamefont {Aime}, \citenamefont
  {Milani}, \citenamefont {Lauricella}, \citenamefont {Succi},\ and\
  \citenamefont {Weitz}}]{tiribocchi2020novel}%
  \BibitemOpen
  \bibfield  {author} {\bibinfo {author} {\bibfnamefont {A.}~\bibnamefont
  {Tiribocchi}}, \bibinfo {author} {\bibfnamefont {A.}~\bibnamefont
  {Montessori}}, \bibinfo {author} {\bibfnamefont {S.}~\bibnamefont {Aime}},
  \bibinfo {author} {\bibfnamefont {M.}~\bibnamefont {Milani}}, \bibinfo
  {author} {\bibfnamefont {M.}~\bibnamefont {Lauricella}}, \bibinfo {author}
  {\bibfnamefont {S.}~\bibnamefont {Succi}}, \ and\ \bibinfo {author}
  {\bibfnamefont {D.}~\bibnamefont {Weitz}},\ }\href@noop {} {\bibfield
  {journal} {\bibinfo  {journal} {Physics of Fluids}\ }\textbf {\bibinfo
  {volume} {32}} (\bibinfo {year} {2020})}\BibitemShut {NoStop}%
\bibitem [{\citenamefont {Chiappini}\ \emph {et~al.}(2019)\citenamefont
  {Chiappini}, \citenamefont {Sbragaglia}, \citenamefont {Xue},\ and\
  \citenamefont {Falcucci}}]{chiappini2019hydrodynamic}%
  \BibitemOpen
  \bibfield  {author} {\bibinfo {author} {\bibfnamefont {D.}~\bibnamefont
  {Chiappini}}, \bibinfo {author} {\bibfnamefont {M.}~\bibnamefont
  {Sbragaglia}}, \bibinfo {author} {\bibfnamefont {X.}~\bibnamefont {Xue}}, \
  and\ \bibinfo {author} {\bibfnamefont {G.}~\bibnamefont {Falcucci}},\
  }\href@noop {} {\bibfield  {journal} {\bibinfo  {journal} {Physical Review
  E}\ }\textbf {\bibinfo {volume} {99}},\ \bibinfo {pages} {053305} (\bibinfo
  {year} {2019})}\BibitemShut {NoStop}%
\bibitem [{\citenamefont {Montessori}, \citenamefont {Lauricella},\ and\
  \citenamefont {Succi}(2019)}]{montessori2019mesoscale}%
  \BibitemOpen
  \bibfield  {author} {\bibinfo {author} {\bibfnamefont {A.}~\bibnamefont
  {Montessori}}, \bibinfo {author} {\bibfnamefont {M.}~\bibnamefont
  {Lauricella}}, \ and\ \bibinfo {author} {\bibfnamefont {S.}~\bibnamefont
  {Succi}},\ }\href@noop {} {\bibfield  {journal} {\bibinfo  {journal}
  {Philosophical Transactions of the Royal Society A}\ }\textbf {\bibinfo
  {volume} {377}},\ \bibinfo {pages} {20180149} (\bibinfo {year}
  {2019})}\BibitemShut {NoStop}%
\bibitem [{\citenamefont {Liu}, \citenamefont {Valocchi},\ and\ \citenamefont
  {Kang}(2012)}]{liu2012three}%
  \BibitemOpen
  \bibfield  {author} {\bibinfo {author} {\bibfnamefont {H.}~\bibnamefont
  {Liu}}, \bibinfo {author} {\bibfnamefont {A.~J.}\ \bibnamefont {Valocchi}}, \
  and\ \bibinfo {author} {\bibfnamefont {Q.}~\bibnamefont {Kang}},\ }\href@noop
  {} {\bibfield  {journal} {\bibinfo  {journal} {Physical Review
  E—Statistical, Nonlinear, and Soft Matter Physics}\ }\textbf {\bibinfo
  {volume} {85}},\ \bibinfo {pages} {046309} (\bibinfo {year}
  {2012})}\BibitemShut {NoStop}%
\bibitem [{\citenamefont {He}, \citenamefont {Chen},\ and\ \citenamefont
  {Zhang}(1999)}]{he1999lattice}%
  \BibitemOpen
  \bibfield  {author} {\bibinfo {author} {\bibfnamefont {X.}~\bibnamefont
  {He}}, \bibinfo {author} {\bibfnamefont {S.}~\bibnamefont {Chen}}, \ and\
  \bibinfo {author} {\bibfnamefont {R.}~\bibnamefont {Zhang}},\ }\href@noop {}
  {\bibfield  {journal} {\bibinfo  {journal} {Journal of computational
  physics}\ }\textbf {\bibinfo {volume} {152}},\ \bibinfo {pages} {642}
  (\bibinfo {year} {1999})}\BibitemShut {NoStop}%
\bibitem [{\citenamefont {Pelusi}\ \emph {et~al.}(2023)\citenamefont {Pelusi},
  \citenamefont {Guglietta}, \citenamefont {Sega}, \citenamefont {Aouane},\
  and\ \citenamefont {Harting}}]{pelusi2023sharp}%
  \BibitemOpen
  \bibfield  {author} {\bibinfo {author} {\bibfnamefont {F.}~\bibnamefont
  {Pelusi}}, \bibinfo {author} {\bibfnamefont {F.}~\bibnamefont {Guglietta}},
  \bibinfo {author} {\bibfnamefont {M.}~\bibnamefont {Sega}}, \bibinfo {author}
  {\bibfnamefont {O.}~\bibnamefont {Aouane}}, \ and\ \bibinfo {author}
  {\bibfnamefont {J.}~\bibnamefont {Harting}},\ }\href@noop {} {\bibfield
  {journal} {\bibinfo  {journal} {Physics of Fluids}\ }\textbf {\bibinfo
  {volume} {35}} (\bibinfo {year} {2023})}\BibitemShut {NoStop}%
\bibitem [{\citenamefont {Guglietta}\ \emph {et~al.}(2023)\citenamefont
  {Guglietta}, \citenamefont {Pelusi}, \citenamefont {Sega}, \citenamefont
  {Aouane},\ and\ \citenamefont {Harting}}]{guglietta2023suspensions}%
  \BibitemOpen
  \bibfield  {author} {\bibinfo {author} {\bibfnamefont {F.}~\bibnamefont
  {Guglietta}}, \bibinfo {author} {\bibfnamefont {F.}~\bibnamefont {Pelusi}},
  \bibinfo {author} {\bibfnamefont {M.}~\bibnamefont {Sega}}, \bibinfo {author}
  {\bibfnamefont {O.}~\bibnamefont {Aouane}}, \ and\ \bibinfo {author}
  {\bibfnamefont {J.}~\bibnamefont {Harting}},\ }\href@noop {} {\bibfield
  {journal} {\bibinfo  {journal} {Journal of Fluid Mechanics}\ }\textbf
  {\bibinfo {volume} {971}},\ \bibinfo {pages} {A13} (\bibinfo {year}
  {2023})}\BibitemShut {NoStop}%
\bibitem [{\citenamefont {Bonaccorso}\ \emph {et~al.}(2020)\citenamefont
  {Bonaccorso}, \citenamefont {Montessori}, \citenamefont {Tiribocchi},
  \citenamefont {Amati}, \citenamefont {Bernaschi}, \citenamefont
  {Lauricella},\ and\ \citenamefont {Succi}}]{bonaccorso2020lbsoft}%
  \BibitemOpen
  \bibfield  {author} {\bibinfo {author} {\bibfnamefont {F.}~\bibnamefont
  {Bonaccorso}}, \bibinfo {author} {\bibfnamefont {A.}~\bibnamefont
  {Montessori}}, \bibinfo {author} {\bibfnamefont {A.}~\bibnamefont
  {Tiribocchi}}, \bibinfo {author} {\bibfnamefont {G.}~\bibnamefont {Amati}},
  \bibinfo {author} {\bibfnamefont {M.}~\bibnamefont {Bernaschi}}, \bibinfo
  {author} {\bibfnamefont {M.}~\bibnamefont {Lauricella}}, \ and\ \bibinfo
  {author} {\bibfnamefont {S.}~\bibnamefont {Succi}},\ }\href@noop {}
  {\bibfield  {journal} {\bibinfo  {journal} {Computer Physics Communications}\
  }\textbf {\bibinfo {volume} {256}},\ \bibinfo {pages} {107455} (\bibinfo
  {year} {2020})}\BibitemShut {NoStop}%
\bibitem [{\citenamefont {Nguyen}\ and\ \citenamefont
  {Ladd}(2002)}]{nguyen2002lubrication}%
  \BibitemOpen
  \bibfield  {author} {\bibinfo {author} {\bibfnamefont {N.-Q.}\ \bibnamefont
  {Nguyen}}\ and\ \bibinfo {author} {\bibfnamefont {A.}~\bibnamefont {Ladd}},\
  }\href@noop {} {\bibfield  {journal} {\bibinfo  {journal} {Physical Review
  E}\ }\textbf {\bibinfo {volume} {66}},\ \bibinfo {pages} {046708} (\bibinfo
  {year} {2002})}\BibitemShut {NoStop}%
\bibitem [{\citenamefont {Ladd}\ and\ \citenamefont
  {Verberg}(2001)}]{ladd2001lattice}%
  \BibitemOpen
  \bibfield  {author} {\bibinfo {author} {\bibfnamefont {A.~J.}\ \bibnamefont
  {Ladd}}\ and\ \bibinfo {author} {\bibfnamefont {R.}~\bibnamefont {Verberg}},\
  }\href@noop {} {\bibfield  {journal} {\bibinfo  {journal} {Journal of
  statistical physics}\ }\textbf {\bibinfo {volume} {104}},\ \bibinfo {pages}
  {1191} (\bibinfo {year} {2001})}\BibitemShut {NoStop}%
\bibitem [{\citenamefont {Monteferrante}\ \emph {et~al.}(2021)\citenamefont
  {Monteferrante}, \citenamefont {Montessori}, \citenamefont {Succi},
  \citenamefont {Pisignano},\ and\ \citenamefont
  {Lauricella}}]{monteferrante2021lattice}%
  \BibitemOpen
  \bibfield  {author} {\bibinfo {author} {\bibfnamefont {M.}~\bibnamefont
  {Monteferrante}}, \bibinfo {author} {\bibfnamefont {A.}~\bibnamefont
  {Montessori}}, \bibinfo {author} {\bibfnamefont {S.}~\bibnamefont {Succi}},
  \bibinfo {author} {\bibfnamefont {D.}~\bibnamefont {Pisignano}}, \ and\
  \bibinfo {author} {\bibfnamefont {M.}~\bibnamefont {Lauricella}},\
  }\href@noop {} {\bibfield  {journal} {\bibinfo  {journal} {Physics of
  Fluids}\ }\textbf {\bibinfo {volume} {33}} (\bibinfo {year}
  {2021})}\BibitemShut {NoStop}%
\bibitem [{\citenamefont {Malaspinas}, \citenamefont {Fi{\'e}tier},\ and\
  \citenamefont {Deville}(2010)}]{malaspinas2010lattice}%
  \BibitemOpen
  \bibfield  {author} {\bibinfo {author} {\bibfnamefont {O.}~\bibnamefont
  {Malaspinas}}, \bibinfo {author} {\bibfnamefont {N.}~\bibnamefont
  {Fi{\'e}tier}}, \ and\ \bibinfo {author} {\bibfnamefont {M.}~\bibnamefont
  {Deville}},\ }\href@noop {} {\bibfield  {journal} {\bibinfo  {journal}
  {Journal of Non-Newtonian Fluid Mechanics}\ }\textbf {\bibinfo {volume}
  {165}},\ \bibinfo {pages} {1637} (\bibinfo {year} {2010})}\BibitemShut
  {NoStop}%
\bibitem [{\citenamefont {Berk~Usta}, \citenamefont {Ladd},\ and\ \citenamefont
  {Butler}(2005)}]{berk2005lattice}%
  \BibitemOpen
  \bibfield  {author} {\bibinfo {author} {\bibfnamefont {O.}~\bibnamefont
  {Berk~Usta}}, \bibinfo {author} {\bibfnamefont {A.~J.}\ \bibnamefont {Ladd}},
  \ and\ \bibinfo {author} {\bibfnamefont {J.~E.}\ \bibnamefont {Butler}},\
  }\href@noop {} {\bibfield  {journal} {\bibinfo  {journal} {The Journal of
  chemical physics}\ }\textbf {\bibinfo {volume} {122}} (\bibinfo {year}
  {2005})}\BibitemShut {NoStop}%
\bibitem [{\citenamefont {Ahlrichs}\ and\ \citenamefont
  {D{\"u}nweg}(1999)}]{ahlrichs1999simulation}%
  \BibitemOpen
  \bibfield  {author} {\bibinfo {author} {\bibfnamefont {P.}~\bibnamefont
  {Ahlrichs}}\ and\ \bibinfo {author} {\bibfnamefont {B.}~\bibnamefont
  {D{\"u}nweg}},\ }\href@noop {} {\bibfield  {journal} {\bibinfo  {journal}
  {The Journal of chemical physics}\ }\textbf {\bibinfo {volume} {111}},\
  \bibinfo {pages} {8225} (\bibinfo {year} {1999})}\BibitemShut {NoStop}%
\bibitem [{\citenamefont {Ahlrichs}\ and\ \citenamefont
  {D{\"u}nweg}(1998)}]{ahlrichs1998lattice}%
  \BibitemOpen
  \bibfield  {author} {\bibinfo {author} {\bibfnamefont {P.}~\bibnamefont
  {Ahlrichs}}\ and\ \bibinfo {author} {\bibfnamefont {B.}~\bibnamefont
  {D{\"u}nweg}},\ }\href@noop {} {\bibfield  {journal} {\bibinfo  {journal}
  {International Journal of Modern Physics C}\ }\textbf {\bibinfo {volume}
  {9}},\ \bibinfo {pages} {1429} (\bibinfo {year} {1998})}\BibitemShut
  {NoStop}%
\bibitem [{\citenamefont {Xiong}\ \emph {et~al.}(2025)\citenamefont {Xiong},
  \citenamefont {Wang}, \citenamefont {Huang},\ and\ \citenamefont
  {Luo}}]{xiong2025thermodynamically}%
  \BibitemOpen
  \bibfield  {author} {\bibinfo {author} {\bibfnamefont {F.}~\bibnamefont
  {Xiong}}, \bibinfo {author} {\bibfnamefont {L.}~\bibnamefont {Wang}},
  \bibinfo {author} {\bibfnamefont {J.}~\bibnamefont {Huang}}, \ and\ \bibinfo
  {author} {\bibfnamefont {K.}~\bibnamefont {Luo}},\ }\href@noop {} {\bibfield
  {journal} {\bibinfo  {journal} {Journal of Scientific Computing}\ }\textbf
  {\bibinfo {volume} {103}},\ \bibinfo {pages} {1} (\bibinfo {year}
  {2025})}\BibitemShut {NoStop}%
\bibitem [{\citenamefont {Liu}\ \emph {et~al.}(2024)\citenamefont {Liu},
  \citenamefont {Chai}, \citenamefont {Shi},\ and\ \citenamefont
  {Yuan}}]{liu2024consistent}%
  \BibitemOpen
  \bibfield  {author} {\bibinfo {author} {\bibfnamefont {X.}~\bibnamefont
  {Liu}}, \bibinfo {author} {\bibfnamefont {Z.}~\bibnamefont {Chai}}, \bibinfo
  {author} {\bibfnamefont {B.}~\bibnamefont {Shi}}, \ and\ \bibinfo {author}
  {\bibfnamefont {X.}~\bibnamefont {Yuan}},\ }\href@noop {} {\bibfield
  {journal} {\bibinfo  {journal} {Physica D: Nonlinear Phenomena}\ }\textbf
  {\bibinfo {volume} {468}},\ \bibinfo {pages} {134294} (\bibinfo {year}
  {2024})}\BibitemShut {NoStop}%
\bibitem [{\citenamefont {Wang}\ \emph {et~al.}(2021)\citenamefont {Wang},
  \citenamefont {Wei}, \citenamefont {Li}, \citenamefont {Chai},\ and\
  \citenamefont {Shi}}]{wang2021lattice}%
  \BibitemOpen
  \bibfield  {author} {\bibinfo {author} {\bibfnamefont {L.}~\bibnamefont
  {Wang}}, \bibinfo {author} {\bibfnamefont {Z.}~\bibnamefont {Wei}}, \bibinfo
  {author} {\bibfnamefont {T.}~\bibnamefont {Li}}, \bibinfo {author}
  {\bibfnamefont {Z.}~\bibnamefont {Chai}}, \ and\ \bibinfo {author}
  {\bibfnamefont {B.}~\bibnamefont {Shi}},\ }\href@noop {} {\bibfield
  {journal} {\bibinfo  {journal} {Applied Mathematical Modelling}\ }\textbf
  {\bibinfo {volume} {95}},\ \bibinfo {pages} {361} (\bibinfo {year}
  {2021})}\BibitemShut {NoStop}%
\bibitem [{\citenamefont {Lauricella}\ \emph {et~al.}(2018)\citenamefont
  {Lauricella}, \citenamefont {Melchionna}, \citenamefont {Montessori},
  \citenamefont {Pisignano}, \citenamefont {Pontrelli},\ and\ \citenamefont
  {Succi}}]{lauricella2018entropic}%
  \BibitemOpen
  \bibfield  {author} {\bibinfo {author} {\bibfnamefont {M.}~\bibnamefont
  {Lauricella}}, \bibinfo {author} {\bibfnamefont {S.}~\bibnamefont
  {Melchionna}}, \bibinfo {author} {\bibfnamefont {A.}~\bibnamefont
  {Montessori}}, \bibinfo {author} {\bibfnamefont {D.}~\bibnamefont
  {Pisignano}}, \bibinfo {author} {\bibfnamefont {G.}~\bibnamefont
  {Pontrelli}}, \ and\ \bibinfo {author} {\bibfnamefont {S.}~\bibnamefont
  {Succi}},\ }\href@noop {} {\bibfield  {journal} {\bibinfo  {journal}
  {Physical Review E}\ }\textbf {\bibinfo {volume} {97}},\ \bibinfo {pages}
  {033308} (\bibinfo {year} {2018})}\BibitemShut {NoStop}%
\bibitem [{\citenamefont {Kupershtokh}\ and\ \citenamefont
  {Medvedev}(2006)}]{kupershtokh2006lattice}%
  \BibitemOpen
  \bibfield  {author} {\bibinfo {author} {\bibfnamefont {A.}~\bibnamefont
  {Kupershtokh}}\ and\ \bibinfo {author} {\bibfnamefont {D.}~\bibnamefont
  {Medvedev}},\ }\href@noop {} {\bibfield  {journal} {\bibinfo  {journal}
  {Journal of electrostatics}\ }\textbf {\bibinfo {volume} {64}},\ \bibinfo
  {pages} {581} (\bibinfo {year} {2006})}\BibitemShut {NoStop}%
\bibitem [{\citenamefont {Sawant}, \citenamefont {Dorschner},\ and\
  \citenamefont {Karlin}(2021)}]{sawant2021lattice}%
  \BibitemOpen
  \bibfield  {author} {\bibinfo {author} {\bibfnamefont {N.}~\bibnamefont
  {Sawant}}, \bibinfo {author} {\bibfnamefont {B.}~\bibnamefont {Dorschner}}, \
  and\ \bibinfo {author} {\bibfnamefont {I.~V.}\ \bibnamefont {Karlin}},\
  }\href@noop {} {\bibfield  {journal} {\bibinfo  {journal} {Philosophical
  Transactions of the Royal Society A}\ }\textbf {\bibinfo {volume} {379}},\
  \bibinfo {pages} {20200402} (\bibinfo {year} {2021})}\BibitemShut {NoStop}%
\bibitem [{\citenamefont {Lin}\ \emph {et~al.}(2017)\citenamefont {Lin},
  \citenamefont {Luo}, \citenamefont {Fei},\ and\ \citenamefont
  {Succi}}]{lin2017multi}%
  \BibitemOpen
  \bibfield  {author} {\bibinfo {author} {\bibfnamefont {C.}~\bibnamefont
  {Lin}}, \bibinfo {author} {\bibfnamefont {K.~H.}\ \bibnamefont {Luo}},
  \bibinfo {author} {\bibfnamefont {L.}~\bibnamefont {Fei}}, \ and\ \bibinfo
  {author} {\bibfnamefont {S.}~\bibnamefont {Succi}},\ }\href@noop {}
  {\bibfield  {journal} {\bibinfo  {journal} {Scientific reports}\ }\textbf
  {\bibinfo {volume} {7}},\ \bibinfo {pages} {14580} (\bibinfo {year}
  {2017})}\BibitemShut {NoStop}%
\bibitem [{\citenamefont {Tiribocchi}\ \emph
  {et~al.}(2023{\natexlab{a}})\citenamefont {Tiribocchi}, \citenamefont
  {Durve}, \citenamefont {Lauricella}, \citenamefont {Montessori},
  \citenamefont {Marenduzzo},\ and\ \citenamefont
  {Succi}}]{tiribocchi2023crucial}%
  \BibitemOpen
  \bibfield  {author} {\bibinfo {author} {\bibfnamefont {A.}~\bibnamefont
  {Tiribocchi}}, \bibinfo {author} {\bibfnamefont {M.}~\bibnamefont {Durve}},
  \bibinfo {author} {\bibfnamefont {M.}~\bibnamefont {Lauricella}}, \bibinfo
  {author} {\bibfnamefont {A.}~\bibnamefont {Montessori}}, \bibinfo {author}
  {\bibfnamefont {D.}~\bibnamefont {Marenduzzo}}, \ and\ \bibinfo {author}
  {\bibfnamefont {S.}~\bibnamefont {Succi}},\ }\href@noop {} {\bibfield
  {journal} {\bibinfo  {journal} {Nature Communications}\ }\textbf {\bibinfo
  {volume} {14}},\ \bibinfo {pages} {1096} (\bibinfo {year}
  {2023}{\natexlab{a}})}\BibitemShut {NoStop}%
\bibitem [{\citenamefont {Carenza}\ \emph {et~al.}(2020)\citenamefont
  {Carenza}, \citenamefont {Gonnella}, \citenamefont {Marenduzzo},\ and\
  \citenamefont {Negro}}]{carenza2020chaotic}%
  \BibitemOpen
  \bibfield  {author} {\bibinfo {author} {\bibfnamefont {L.~N.}\ \bibnamefont
  {Carenza}}, \bibinfo {author} {\bibfnamefont {G.}~\bibnamefont {Gonnella}},
  \bibinfo {author} {\bibfnamefont {D.}~\bibnamefont {Marenduzzo}}, \ and\
  \bibinfo {author} {\bibfnamefont {G.}~\bibnamefont {Negro}},\ }\href@noop {}
  {\bibfield  {journal} {\bibinfo  {journal} {Physica A: Statistical Mechanics
  and its Applications}\ }\textbf {\bibinfo {volume} {559}},\ \bibinfo {pages}
  {125025} (\bibinfo {year} {2020})}\BibitemShut {NoStop}%
\bibitem [{\citenamefont {de~Graaf}\ \emph {et~al.}(2016)\citenamefont
  {de~Graaf}, \citenamefont {Menke}, \citenamefont {Mathijssen}, \citenamefont
  {Fabritius}, \citenamefont {Holm},\ and\ \citenamefont
  {Shendruk}}]{de2016lattice}%
  \BibitemOpen
  \bibfield  {author} {\bibinfo {author} {\bibfnamefont {J.}~\bibnamefont
  {de~Graaf}}, \bibinfo {author} {\bibfnamefont {H.}~\bibnamefont {Menke}},
  \bibinfo {author} {\bibfnamefont {A.~J.}\ \bibnamefont {Mathijssen}},
  \bibinfo {author} {\bibfnamefont {M.}~\bibnamefont {Fabritius}}, \bibinfo
  {author} {\bibfnamefont {C.}~\bibnamefont {Holm}}, \ and\ \bibinfo {author}
  {\bibfnamefont {T.~N.}\ \bibnamefont {Shendruk}},\ }\href@noop {} {\bibfield
  {journal} {\bibinfo  {journal} {The Journal of chemical physics}\ }\textbf
  {\bibinfo {volume} {144}} (\bibinfo {year} {2016})}\BibitemShut {NoStop}%
\bibitem [{\citenamefont {Marenduzzo}\ \emph {et~al.}(2007)\citenamefont
  {Marenduzzo}, \citenamefont {Orlandini}, \citenamefont {Cates},\ and\
  \citenamefont {Yeomans}}]{marenduzzo2007steady}%
  \BibitemOpen
  \bibfield  {author} {\bibinfo {author} {\bibfnamefont {D.}~\bibnamefont
  {Marenduzzo}}, \bibinfo {author} {\bibfnamefont {E.}~\bibnamefont
  {Orlandini}}, \bibinfo {author} {\bibfnamefont {M.}~\bibnamefont {Cates}}, \
  and\ \bibinfo {author} {\bibfnamefont {J.}~\bibnamefont {Yeomans}},\
  }\href@noop {} {\bibfield  {journal} {\bibinfo  {journal} {Physical Review
  E—Statistical, Nonlinear, and Soft Matter Physics}\ }\textbf {\bibinfo
  {volume} {76}},\ \bibinfo {pages} {031921} (\bibinfo {year}
  {2007})}\BibitemShut {NoStop}%
\bibitem [{\citenamefont {Lauricella}\ \emph {et~al.}(2025)\citenamefont
  {Lauricella}, \citenamefont {Mukherjee}, \citenamefont {Brandt},
  \citenamefont {Succi}, \citenamefont {Izbassarov},\ and\ \citenamefont
  {Montessori}}]{lauricella2025acclb}%
  \BibitemOpen
  \bibfield  {author} {\bibinfo {author} {\bibfnamefont {M.}~\bibnamefont
  {Lauricella}}, \bibinfo {author} {\bibfnamefont {A.}~\bibnamefont
  {Mukherjee}}, \bibinfo {author} {\bibfnamefont {L.}~\bibnamefont {Brandt}},
  \bibinfo {author} {\bibfnamefont {S.}~\bibnamefont {Succi}}, \bibinfo
  {author} {\bibfnamefont {D.}~\bibnamefont {Izbassarov}}, \ and\ \bibinfo
  {author} {\bibfnamefont {A.}~\bibnamefont {Montessori}},\ }\href@noop {}
  {\bibfield  {journal} {\bibinfo  {journal} {arXiv preprint arXiv:2505.01126}\
  } (\bibinfo {year} {2025})}\BibitemShut {NoStop}%
\bibitem [{\citenamefont {Latt}\ and\ \citenamefont
  {Coreixas}(2025)}]{latt2025multi}%
  \BibitemOpen
  \bibfield  {author} {\bibinfo {author} {\bibfnamefont {J.}~\bibnamefont
  {Latt}}\ and\ \bibinfo {author} {\bibfnamefont {C.}~\bibnamefont
  {Coreixas}},\ }\href@noop {} {\bibfield  {journal} {\bibinfo  {journal}
  {arXiv preprint arXiv:2506.09242}\ } (\bibinfo {year} {2025})}\BibitemShut
  {NoStop}%
\bibitem [{\citenamefont {Tiribocchi}\ \emph
  {et~al.}(2023{\natexlab{b}})\citenamefont {Tiribocchi}, \citenamefont
  {Montessori}, \citenamefont {Amati}, \citenamefont {Bernaschi}, \citenamefont
  {Bonaccorso}, \citenamefont {Orlandini}, \citenamefont {Succi},\ and\
  \citenamefont {Lauricella}}]{tiribocchi2023lightweight}%
  \BibitemOpen
  \bibfield  {author} {\bibinfo {author} {\bibfnamefont {A.}~\bibnamefont
  {Tiribocchi}}, \bibinfo {author} {\bibfnamefont {A.}~\bibnamefont
  {Montessori}}, \bibinfo {author} {\bibfnamefont {G.}~\bibnamefont {Amati}},
  \bibinfo {author} {\bibfnamefont {M.}~\bibnamefont {Bernaschi}}, \bibinfo
  {author} {\bibfnamefont {F.}~\bibnamefont {Bonaccorso}}, \bibinfo {author}
  {\bibfnamefont {S.}~\bibnamefont {Orlandini}}, \bibinfo {author}
  {\bibfnamefont {S.}~\bibnamefont {Succi}}, \ and\ \bibinfo {author}
  {\bibfnamefont {M.}~\bibnamefont {Lauricella}},\ }\href@noop {} {\bibfield
  {journal} {\bibinfo  {journal} {The Journal of Chemical Physics}\ }\textbf
  {\bibinfo {volume} {158}} (\bibinfo {year} {2023}{\natexlab{b}})}\BibitemShut
  {NoStop}%
\bibitem [{\citenamefont {Patel}\ \emph {et~al.}(2023)\citenamefont {Patel},
  \citenamefont {Zhao}, \citenamefont {Lee},\ and\ \citenamefont
  {Balakrishnan}}]{patel2023imexlbm}%
  \BibitemOpen
  \bibfield  {author} {\bibinfo {author} {\bibfnamefont {S.}~\bibnamefont
  {Patel}}, \bibinfo {author} {\bibfnamefont {C.}~\bibnamefont {Zhao}},
  \bibinfo {author} {\bibfnamefont {T.}~\bibnamefont {Lee}}, \ and\ \bibinfo
  {author} {\bibfnamefont {R.}~\bibnamefont {Balakrishnan}},\ }\href@noop {}
  {\bibfield  {journal} {\bibinfo  {journal} {Bulletin of the American Physical
  Society}\ } (\bibinfo {year} {2023})}\BibitemShut {NoStop}%
\bibitem [{\citenamefont {Zhao}\ \emph {et~al.}(2023)\citenamefont {Zhao},
  \citenamefont {Wang}, \citenamefont {Zhang}, \citenamefont {Zhou},
  \citenamefont {Chen}, \citenamefont {Viswanathan},\ and\ \citenamefont
  {Kang}}]{zhao2023minireview}%
  \BibitemOpen
  \bibfield  {author} {\bibinfo {author} {\bibfnamefont {J.}~\bibnamefont
  {Zhao}}, \bibinfo {author} {\bibfnamefont {J.}~\bibnamefont {Wang}}, \bibinfo
  {author} {\bibfnamefont {G.}~\bibnamefont {Zhang}}, \bibinfo {author}
  {\bibfnamefont {D.}~\bibnamefont {Zhou}}, \bibinfo {author} {\bibfnamefont
  {L.}~\bibnamefont {Chen}}, \bibinfo {author} {\bibfnamefont {H.}~\bibnamefont
  {Viswanathan}}, \ and\ \bibinfo {author} {\bibfnamefont {Q.}~\bibnamefont
  {Kang}},\ }\href@noop {} {\bibfield  {journal} {\bibinfo  {journal} {Energy
  \& Fuels}\ }\textbf {\bibinfo {volume} {37}},\ \bibinfo {pages} {1511}
  (\bibinfo {year} {2023})}\BibitemShut {NoStop}%
\bibitem [{\citenamefont {Forster}(2018)}]{forster2018hydrodynamic}%
  \BibitemOpen
  \bibfield  {author} {\bibinfo {author} {\bibfnamefont {D.}~\bibnamefont
  {Forster}},\ }\href@noop {} {\emph {\bibinfo {title} {Hydrodynamic
  fluctuations, broken symmetry, and correlation functions}}}\ (\bibinfo
  {publisher} {CRC Press},\ \bibinfo {year} {2018})\BibitemShut {NoStop}%
\bibitem [{\citenamefont {Zwanzig}(2001)}]{zwanzig2001nonequilibrium}%
  \BibitemOpen
  \bibfield  {author} {\bibinfo {author} {\bibfnamefont {R.}~\bibnamefont
  {Zwanzig}},\ }\href@noop {} {\emph {\bibinfo {title} {Nonequilibrium
  statistical mechanics}}}\ (\bibinfo  {publisher} {Oxford university press},\
  \bibinfo {year} {2001})\BibitemShut {NoStop}%
\bibitem [{\citenamefont {Landau}\ and\ \citenamefont
  {Lifshitz}(1987)}]{landau1987fluid}%
  \BibitemOpen
  \bibfield  {author} {\bibinfo {author} {\bibfnamefont {L.~D.}\ \bibnamefont
  {Landau}}\ and\ \bibinfo {author} {\bibfnamefont {E.~M.}\ \bibnamefont
  {Lifshitz}},\ }\href@noop {} {\emph {\bibinfo {title} {Fluid Mechanics:
  Volume 6}}},\ Vol.~\bibinfo {volume} {6}\ (\bibinfo  {publisher} {Elsevier},\
  \bibinfo {year} {1987})\BibitemShut {NoStop}%
\bibitem [{\citenamefont {Frenkel}\ and\ \citenamefont
  {Smit}(2023)}]{frenkel2023understanding}%
  \BibitemOpen
  \bibfield  {author} {\bibinfo {author} {\bibfnamefont {D.}~\bibnamefont
  {Frenkel}}\ and\ \bibinfo {author} {\bibfnamefont {B.}~\bibnamefont {Smit}},\
  }\href@noop {} {\emph {\bibinfo {title} {Understanding molecular simulation:
  from algorithms to applications}}}\ (\bibinfo  {publisher} {Elsevier},\
  \bibinfo {year} {2023})\BibitemShut {NoStop}%
\bibitem [{\citenamefont {Bussi}\ and\ \citenamefont
  {Parrinello}(2008)}]{bussi2008stochastic}%
  \BibitemOpen
  \bibfield  {author} {\bibinfo {author} {\bibfnamefont {G.}~\bibnamefont
  {Bussi}}\ and\ \bibinfo {author} {\bibfnamefont {M.}~\bibnamefont
  {Parrinello}},\ }\href@noop {} {\bibfield  {journal} {\bibinfo  {journal}
  {Computer Physics Communications}\ }\textbf {\bibinfo {volume} {179}},\
  \bibinfo {pages} {26} (\bibinfo {year} {2008})}\BibitemShut {NoStop}%
\bibitem [{\citenamefont {Bussi}, \citenamefont {Donadio},\ and\ \citenamefont
  {Parrinello}(2007)}]{bussi2007canonical}%
  \BibitemOpen
  \bibfield  {author} {\bibinfo {author} {\bibfnamefont {G.}~\bibnamefont
  {Bussi}}, \bibinfo {author} {\bibfnamefont {D.}~\bibnamefont {Donadio}}, \
  and\ \bibinfo {author} {\bibfnamefont {M.}~\bibnamefont {Parrinello}},\
  }\href@noop {} {\bibfield  {journal} {\bibinfo  {journal} {The Journal of
  chemical physics}\ }\textbf {\bibinfo {volume} {126}} (\bibinfo {year}
  {2007})}\BibitemShut {NoStop}%
\bibitem [{\citenamefont {Tanygin}\ and\ \citenamefont
  {Melchionna}(2024)}]{tanygin2024comparison}%
  \BibitemOpen
  \bibfield  {author} {\bibinfo {author} {\bibfnamefont {B.}~\bibnamefont
  {Tanygin}}\ and\ \bibinfo {author} {\bibfnamefont {S.}~\bibnamefont
  {Melchionna}},\ }\href@noop {} {\bibfield  {journal} {\bibinfo  {journal}
  {Computer Physics Communications}\ }\textbf {\bibinfo {volume} {299}},\
  \bibinfo {pages} {109152} (\bibinfo {year} {2024})}\BibitemShut {NoStop}%
\bibitem [{\citenamefont {Bussi}\ and\ \citenamefont
  {Parrinello}(2007)}]{bussi2007accurate}%
  \BibitemOpen
  \bibfield  {author} {\bibinfo {author} {\bibfnamefont {G.}~\bibnamefont
  {Bussi}}\ and\ \bibinfo {author} {\bibfnamefont {M.}~\bibnamefont
  {Parrinello}},\ }\href@noop {} {\bibfield  {journal} {\bibinfo  {journal}
  {Physical Review E—Statistical, Nonlinear, and Soft Matter Physics}\
  }\textbf {\bibinfo {volume} {75}},\ \bibinfo {pages} {056707} (\bibinfo
  {year} {2007})}\BibitemShut {NoStop}%
\bibitem [{\citenamefont {Martyna}, \citenamefont {Klein},\ and\ \citenamefont
  {Tuckerman}(1992)}]{martyna1992nose}%
  \BibitemOpen
  \bibfield  {author} {\bibinfo {author} {\bibfnamefont {G.~J.}\ \bibnamefont
  {Martyna}}, \bibinfo {author} {\bibfnamefont {M.~L.}\ \bibnamefont {Klein}},
  \ and\ \bibinfo {author} {\bibfnamefont {M.}~\bibnamefont {Tuckerman}},\
  }\href@noop {} {\bibfield  {journal} {\bibinfo  {journal} {The Journal of
  chemical physics}\ }\textbf {\bibinfo {volume} {97}},\ \bibinfo {pages}
  {2635} (\bibinfo {year} {1992})}\BibitemShut {NoStop}%
\bibitem [{\citenamefont {Ladd}(1993)}]{ladd1993short}%
  \BibitemOpen
  \bibfield  {author} {\bibinfo {author} {\bibfnamefont {A.~J.}\ \bibnamefont
  {Ladd}},\ }\href@noop {} {\bibfield  {journal} {\bibinfo  {journal} {Physical
  Review Letters}\ }\textbf {\bibinfo {volume} {70}},\ \bibinfo {pages} {1339}
  (\bibinfo {year} {1993})}\BibitemShut {NoStop}%
\bibitem [{\citenamefont {Ladd}(1994)}]{ladd1994numerical}%
  \BibitemOpen
  \bibfield  {author} {\bibinfo {author} {\bibfnamefont {A.~J.}\ \bibnamefont
  {Ladd}},\ }\href@noop {} {\bibfield  {journal} {\bibinfo  {journal} {Journal
  of fluid mechanics}\ }\textbf {\bibinfo {volume} {271}},\ \bibinfo {pages}
  {285} (\bibinfo {year} {1994})}\BibitemShut {NoStop}%
\bibitem [{\citenamefont {D{\"u}nweg}, \citenamefont {Schiller},\ and\
  \citenamefont {Ladd}(2007)}]{dunweg2007statistical}%
  \BibitemOpen
  \bibfield  {author} {\bibinfo {author} {\bibfnamefont {B.}~\bibnamefont
  {D{\"u}nweg}}, \bibinfo {author} {\bibfnamefont {U.~D.}\ \bibnamefont
  {Schiller}}, \ and\ \bibinfo {author} {\bibfnamefont {A.~J.}\ \bibnamefont
  {Ladd}},\ }\href@noop {} {\bibfield  {journal} {\bibinfo  {journal} {Physical
  Review E—Statistical, Nonlinear, and Soft Matter Physics}\ }\textbf
  {\bibinfo {volume} {76}},\ \bibinfo {pages} {036704} (\bibinfo {year}
  {2007})}\BibitemShut {NoStop}%
\bibitem [{\citenamefont {Adhikari}\ \emph {et~al.}(2005)\citenamefont
  {Adhikari}, \citenamefont {Stratford}, \citenamefont {Cates},\ and\
  \citenamefont {Wagner}}]{adhikari2005fluctuating}%
  \BibitemOpen
  \bibfield  {author} {\bibinfo {author} {\bibfnamefont {R.}~\bibnamefont
  {Adhikari}}, \bibinfo {author} {\bibfnamefont {K.}~\bibnamefont {Stratford}},
  \bibinfo {author} {\bibfnamefont {M.}~\bibnamefont {Cates}}, \ and\ \bibinfo
  {author} {\bibfnamefont {A.}~\bibnamefont {Wagner}},\ }\href@noop {}
  {\bibfield  {journal} {\bibinfo  {journal} {Europhysics Letters}\ }\textbf
  {\bibinfo {volume} {71}},\ \bibinfo {pages} {473} (\bibinfo {year}
  {2005})}\BibitemShut {NoStop}%
\bibitem [{\citenamefont {Gross}\ \emph {et~al.}(2011)\citenamefont {Gross},
  \citenamefont {Adhikari}, \citenamefont {Cates},\ and\ \citenamefont
  {Varnik}}]{gross2011modelling}%
  \BibitemOpen
  \bibfield  {author} {\bibinfo {author} {\bibfnamefont {M.}~\bibnamefont
  {Gross}}, \bibinfo {author} {\bibfnamefont {R.}~\bibnamefont {Adhikari}},
  \bibinfo {author} {\bibfnamefont {M.}~\bibnamefont {Cates}}, \ and\ \bibinfo
  {author} {\bibfnamefont {F.}~\bibnamefont {Varnik}},\ }\href@noop {}
  {\bibfield  {journal} {\bibinfo  {journal} {Philosophical Transactions of the
  Royal Society A: Mathematical, Physical and Engineering Sciences}\ }\textbf
  {\bibinfo {volume} {369}},\ \bibinfo {pages} {2274} (\bibinfo {year}
  {2011})}\BibitemShut {NoStop}%
\bibitem [{\citenamefont {Belardinelli}\ \emph {et~al.}(2015)\citenamefont
  {Belardinelli}, \citenamefont {Sbragaglia}, \citenamefont {Biferale},
  \citenamefont {Gross},\ and\ \citenamefont
  {Varnik}}]{belardinelli2015fluctuating}%
  \BibitemOpen
  \bibfield  {author} {\bibinfo {author} {\bibfnamefont {D.}~\bibnamefont
  {Belardinelli}}, \bibinfo {author} {\bibfnamefont {M.}~\bibnamefont
  {Sbragaglia}}, \bibinfo {author} {\bibfnamefont {L.}~\bibnamefont
  {Biferale}}, \bibinfo {author} {\bibfnamefont {M.}~\bibnamefont {Gross}}, \
  and\ \bibinfo {author} {\bibfnamefont {F.}~\bibnamefont {Varnik}},\
  }\href@noop {} {\bibfield  {journal} {\bibinfo  {journal} {Physical Review
  E}\ }\textbf {\bibinfo {volume} {91}},\ \bibinfo {pages} {023313} (\bibinfo
  {year} {2015})}\BibitemShut {NoStop}%
\bibitem [{\citenamefont {Sbragaglia}\ and\ \citenamefont
  {Belardinelli}(2013)}]{sbragaglia2013interaction}%
  \BibitemOpen
  \bibfield  {author} {\bibinfo {author} {\bibfnamefont {M.}~\bibnamefont
  {Sbragaglia}}\ and\ \bibinfo {author} {\bibfnamefont {D.}~\bibnamefont
  {Belardinelli}},\ }\href@noop {} {\bibfield  {journal} {\bibinfo  {journal}
  {Physical Review E—Statistical, Nonlinear, and Soft Matter Physics}\
  }\textbf {\bibinfo {volume} {88}},\ \bibinfo {pages} {013306} (\bibinfo
  {year} {2013})}\BibitemShut {NoStop}%
\bibitem [{\citenamefont {Belardinelli}\ \emph {et~al.}(2019)\citenamefont
  {Belardinelli}, \citenamefont {Sbragaglia}, \citenamefont {Benzi},\ and\
  \citenamefont {Ciliberto}}]{belardinelli2019lattice}%
  \BibitemOpen
  \bibfield  {author} {\bibinfo {author} {\bibfnamefont {D.}~\bibnamefont
  {Belardinelli}}, \bibinfo {author} {\bibfnamefont {M.}~\bibnamefont
  {Sbragaglia}}, \bibinfo {author} {\bibfnamefont {R.}~\bibnamefont {Benzi}}, \
  and\ \bibinfo {author} {\bibfnamefont {S.}~\bibnamefont {Ciliberto}},\
  }\href@noop {} {\bibfield  {journal} {\bibinfo  {journal} {Physical Review
  E}\ }\textbf {\bibinfo {volume} {99}},\ \bibinfo {pages} {063302} (\bibinfo
  {year} {2019})}\BibitemShut {NoStop}%
\bibitem [{\citenamefont {Wagner}\ and\ \citenamefont
  {Strand}(2016)}]{wagner2016fluctuating}%
  \BibitemOpen
  \bibfield  {author} {\bibinfo {author} {\bibfnamefont {A.~J.}\ \bibnamefont
  {Wagner}}\ and\ \bibinfo {author} {\bibfnamefont {K.}~\bibnamefont
  {Strand}},\ }\href@noop {} {\bibfield  {journal} {\bibinfo  {journal}
  {Physical Review E}\ }\textbf {\bibinfo {volume} {94}},\ \bibinfo {pages}
  {033302} (\bibinfo {year} {2016})}\BibitemShut {NoStop}%
\bibitem [{\citenamefont {Kaehler}\ and\ \citenamefont
  {Wagner}(2013)}]{kaehler2013fluctuating}%
  \BibitemOpen
  \bibfield  {author} {\bibinfo {author} {\bibfnamefont {G.}~\bibnamefont
  {Kaehler}}\ and\ \bibinfo {author} {\bibfnamefont {A.}~\bibnamefont
  {Wagner}},\ }\href@noop {} {\bibfield  {journal} {\bibinfo  {journal}
  {Physical Review E—Statistical, Nonlinear, and Soft Matter Physics}\
  }\textbf {\bibinfo {volume} {87}},\ \bibinfo {pages} {063310} (\bibinfo
  {year} {2013})}\BibitemShut {NoStop}%
\bibitem [{\citenamefont {Xue}\ \emph {et~al.}(2021)\citenamefont {Xue},
  \citenamefont {Biferale}, \citenamefont {Sbragaglia},\ and\ \citenamefont
  {Toschi}}]{xue2021lattice}%
  \BibitemOpen
  \bibfield  {author} {\bibinfo {author} {\bibfnamefont {X.}~\bibnamefont
  {Xue}}, \bibinfo {author} {\bibfnamefont {L.}~\bibnamefont {Biferale}},
  \bibinfo {author} {\bibfnamefont {M.}~\bibnamefont {Sbragaglia}}, \ and\
  \bibinfo {author} {\bibfnamefont {F.}~\bibnamefont {Toschi}},\ }\href@noop {}
  {\bibfield  {journal} {\bibinfo  {journal} {The European Physical Journal E}\
  }\textbf {\bibinfo {volume} {44}},\ \bibinfo {pages} {1} (\bibinfo {year}
  {2021})}\BibitemShut {NoStop}%
\bibitem [{\citenamefont {Malaspinas}(2015)}]{malaspinas2015increasing}%
  \BibitemOpen
  \bibfield  {author} {\bibinfo {author} {\bibfnamefont {O.}~\bibnamefont
  {Malaspinas}},\ }\href@noop {} {\bibfield  {journal} {\bibinfo  {journal}
  {arXiv preprint arXiv:1505.06900}\ } (\bibinfo {year} {2015})}\BibitemShut
  {NoStop}%
\bibitem [{\citenamefont {Coreixas}\ \emph {et~al.}(2017)\citenamefont
  {Coreixas}, \citenamefont {Wissocq}, \citenamefont {Puigt}, \citenamefont
  {Boussuge},\ and\ \citenamefont {Sagaut}}]{coreixas2017recursive}%
  \BibitemOpen
  \bibfield  {author} {\bibinfo {author} {\bibfnamefont {C.}~\bibnamefont
  {Coreixas}}, \bibinfo {author} {\bibfnamefont {G.}~\bibnamefont {Wissocq}},
  \bibinfo {author} {\bibfnamefont {G.}~\bibnamefont {Puigt}}, \bibinfo
  {author} {\bibfnamefont {J.-F.}\ \bibnamefont {Boussuge}}, \ and\ \bibinfo
  {author} {\bibfnamefont {P.}~\bibnamefont {Sagaut}},\ }\href@noop {}
  {\bibfield  {journal} {\bibinfo  {journal} {Physical Review E}\ }\textbf
  {\bibinfo {volume} {96}},\ \bibinfo {pages} {033306} (\bibinfo {year}
  {2017})}\BibitemShut {NoStop}%
\bibitem [{\citenamefont {Mattila}, \citenamefont {Philippi},\ and\
  \citenamefont {Hegele}(2017)}]{mattila2017high}%
  \BibitemOpen
  \bibfield  {author} {\bibinfo {author} {\bibfnamefont {K.~K.}\ \bibnamefont
  {Mattila}}, \bibinfo {author} {\bibfnamefont {P.~C.}\ \bibnamefont
  {Philippi}}, \ and\ \bibinfo {author} {\bibfnamefont {L.~A.}\ \bibnamefont
  {Hegele}},\ }\href@noop {} {\bibfield  {journal} {\bibinfo  {journal}
  {Physics of Fluids}\ }\textbf {\bibinfo {volume} {29}} (\bibinfo {year}
  {2017})}\BibitemShut {NoStop}%
\bibitem [{\citenamefont {Latt}\ and\ \citenamefont
  {Chopard}(2006)}]{latt2006lattice}%
  \BibitemOpen
  \bibfield  {author} {\bibinfo {author} {\bibfnamefont {J.}~\bibnamefont
  {Latt}}\ and\ \bibinfo {author} {\bibfnamefont {B.}~\bibnamefont {Chopard}},\
  }\href@noop {} {\bibfield  {journal} {\bibinfo  {journal} {Mathematics and
  Computers in Simulation}\ }\textbf {\bibinfo {volume} {72}},\ \bibinfo
  {pages} {165} (\bibinfo {year} {2006})}\BibitemShut {NoStop}%
\bibitem [{\citenamefont {Schiller}(2008)}]{schiller2008thermal}%
  \BibitemOpen
  \bibfield  {author} {\bibinfo {author} {\bibfnamefont {U.~D.}\ \bibnamefont
  {Schiller}},\ }\emph {\bibinfo {title} {Thermal fluctuations and boundary
  conditions in the lattice Boltzmann method}},\ \href@noop {} {Ph.D. thesis},\
  \bibinfo  {school} {Johannes Gutenberg Universit{\"a}t Mainz} (\bibinfo
  {year} {2008})\BibitemShut {NoStop}%
\bibitem [{\citenamefont {d'Humi{\`e}res}(2002)}]{d2002multiple}%
  \BibitemOpen
  \bibfield  {author} {\bibinfo {author} {\bibfnamefont {D.}~\bibnamefont
  {d'Humi{\`e}res}},\ }\href@noop {} {\bibfield  {journal} {\bibinfo  {journal}
  {Philosophical Transactions of the Royal Society of London. Series A:
  Mathematical, Physical and Engineering Sciences}\ }\textbf {\bibinfo {volume}
  {360}},\ \bibinfo {pages} {437} (\bibinfo {year} {2002})}\BibitemShut
  {NoStop}%
\bibitem [{\citenamefont {Lallemand}\ and\ \citenamefont
  {Luo}(2000)}]{lallemand2000theory}%
  \BibitemOpen
  \bibfield  {author} {\bibinfo {author} {\bibfnamefont {P.}~\bibnamefont
  {Lallemand}}\ and\ \bibinfo {author} {\bibfnamefont {L.-S.}\ \bibnamefont
  {Luo}},\ }\href@noop {} {\bibfield  {journal} {\bibinfo  {journal} {Physical
  review E}\ }\textbf {\bibinfo {volume} {61}},\ \bibinfo {pages} {6546}
  (\bibinfo {year} {2000})}\BibitemShut {NoStop}%
\bibitem [{\citenamefont {Zhang}, \citenamefont {Shan},\ and\ \citenamefont
  {Chen}(2006)}]{zhang2006efficient}%
  \BibitemOpen
  \bibfield  {author} {\bibinfo {author} {\bibfnamefont {R.}~\bibnamefont
  {Zhang}}, \bibinfo {author} {\bibfnamefont {X.}~\bibnamefont {Shan}}, \ and\
  \bibinfo {author} {\bibfnamefont {H.}~\bibnamefont {Chen}},\ }\href@noop {}
  {\bibfield  {journal} {\bibinfo  {journal} {Physical Review E}\ }\textbf
  {\bibinfo {volume} {74}},\ \bibinfo {pages} {046703} (\bibinfo {year}
  {2006})}\BibitemShut {NoStop}%
\bibitem [{\citenamefont {Chapman}\ and\ \citenamefont
  {Cowling}(1990)}]{chapman1990mathematical}%
  \BibitemOpen
  \bibfield  {author} {\bibinfo {author} {\bibfnamefont {S.}~\bibnamefont
  {Chapman}}\ and\ \bibinfo {author} {\bibfnamefont {T.~G.}\ \bibnamefont
  {Cowling}},\ }\href@noop {} {\emph {\bibinfo {title} {The mathematical theory
  of non-uniform gases: an account of the kinetic theory of viscosity, thermal
  conduction and diffusion in gases}}}\ (\bibinfo  {publisher} {Cambridge
  university press},\ \bibinfo {year} {1990})\BibitemShut {NoStop}%
\bibitem [{\citenamefont {Latt}\ \emph {et~al.}(2008)\citenamefont {Latt},
  \citenamefont {Chopard}, \citenamefont {Malaspinas}, \citenamefont
  {Deville},\ and\ \citenamefont {Michler}}]{latt2008straight}%
  \BibitemOpen
  \bibfield  {author} {\bibinfo {author} {\bibfnamefont {J.}~\bibnamefont
  {Latt}}, \bibinfo {author} {\bibfnamefont {B.}~\bibnamefont {Chopard}},
  \bibinfo {author} {\bibfnamefont {O.}~\bibnamefont {Malaspinas}}, \bibinfo
  {author} {\bibfnamefont {M.}~\bibnamefont {Deville}}, \ and\ \bibinfo
  {author} {\bibfnamefont {A.}~\bibnamefont {Michler}},\ }\href@noop {}
  {\bibfield  {journal} {\bibinfo  {journal} {Physical Review E}\ }\textbf
  {\bibinfo {volume} {77}},\ \bibinfo {pages} {056703} (\bibinfo {year}
  {2008})}\BibitemShut {NoStop}%
\bibitem [{\citenamefont {Shan}, \citenamefont {Yuan},\ and\ \citenamefont
  {Chen}(2006)}]{shan2006kinetic}%
  \BibitemOpen
  \bibfield  {author} {\bibinfo {author} {\bibfnamefont {X.}~\bibnamefont
  {Shan}}, \bibinfo {author} {\bibfnamefont {X.-F.}\ \bibnamefont {Yuan}}, \
  and\ \bibinfo {author} {\bibfnamefont {H.}~\bibnamefont {Chen}},\ }\href@noop
  {} {\bibfield  {journal} {\bibinfo  {journal} {Journal of Fluid Mechanics}\
  }\textbf {\bibinfo {volume} {550}},\ \bibinfo {pages} {413} (\bibinfo {year}
  {2006})}\BibitemShut {NoStop}%
\bibitem [{\citenamefont {Grad}(1949{\natexlab{a}})}]{grad1949note}%
  \BibitemOpen
  \bibfield  {author} {\bibinfo {author} {\bibfnamefont {H.}~\bibnamefont
  {Grad}},\ }\href@noop {} {\bibfield  {journal} {\bibinfo  {journal}
  {Communications on Pure and Applied Mathematics}\ }\textbf {\bibinfo {volume}
  {2}},\ \bibinfo {pages} {325} (\bibinfo {year}
  {1949}{\natexlab{a}})}\BibitemShut {NoStop}%
\bibitem [{\citenamefont {Grad}(1949{\natexlab{b}})}]{grad1949kinetic}%
  \BibitemOpen
  \bibfield  {author} {\bibinfo {author} {\bibfnamefont {H.}~\bibnamefont
  {Grad}},\ }\href@noop {} {\bibfield  {journal} {\bibinfo  {journal}
  {Communications on pure and applied mathematics}\ }\textbf {\bibinfo {volume}
  {2}},\ \bibinfo {pages} {331} (\bibinfo {year}
  {1949}{\natexlab{b}})}\BibitemShut {NoStop}%
\bibitem [{\citenamefont {Seekins}\ and\ \citenamefont
  {Wagner}(2025)}]{seekins2025integer}%
  \BibitemOpen
  \bibfield  {author} {\bibinfo {author} {\bibfnamefont {N.}~\bibnamefont
  {Seekins}}\ and\ \bibinfo {author} {\bibfnamefont {A.}~\bibnamefont
  {Wagner}},\ }\href@noop {} {\bibfield  {journal} {\bibinfo  {journal} {arXiv
  preprint arXiv:2504.18659}\ } (\bibinfo {year} {2025})}\BibitemShut {NoStop}%
\bibitem [{\citenamefont {Blommel}\ and\ \citenamefont
  {Wagner}(2018)}]{blommel2018integer}%
  \BibitemOpen
  \bibfield  {author} {\bibinfo {author} {\bibfnamefont {T.}~\bibnamefont
  {Blommel}}\ and\ \bibinfo {author} {\bibfnamefont {A.~J.}\ \bibnamefont
  {Wagner}},\ }\href@noop {} {\bibfield  {journal} {\bibinfo  {journal}
  {Physical Review E}\ }\textbf {\bibinfo {volume} {97}},\ \bibinfo {pages}
  {023310} (\bibinfo {year} {2018})}\BibitemShut {NoStop}%
\bibitem [{\citenamefont {Ollila}\ \emph {et~al.}(2011)\citenamefont {Ollila},
  \citenamefont {Denniston}, \citenamefont {Karttunen},\ and\ \citenamefont
  {Ala-Nissila}}]{ollila2011fluctuating}%
  \BibitemOpen
  \bibfield  {author} {\bibinfo {author} {\bibfnamefont {S.~T.}\ \bibnamefont
  {Ollila}}, \bibinfo {author} {\bibfnamefont {C.}~\bibnamefont {Denniston}},
  \bibinfo {author} {\bibfnamefont {M.}~\bibnamefont {Karttunen}}, \ and\
  \bibinfo {author} {\bibfnamefont {T.}~\bibnamefont {Ala-Nissila}},\
  }\href@noop {} {\bibfield  {journal} {\bibinfo  {journal} {The Journal of
  chemical physics}\ }\textbf {\bibinfo {volume} {134}} (\bibinfo {year}
  {2011})}\BibitemShut {NoStop}%
\bibitem [{\citenamefont {Bernaschi}\ \emph {et~al.}(2009)\citenamefont
  {Bernaschi}, \citenamefont {Melchionna}, \citenamefont {Succi}, \citenamefont
  {Fyta}, \citenamefont {Kaxiras},\ and\ \citenamefont
  {Sircar}}]{bernaschi2009muphy}%
  \BibitemOpen
  \bibfield  {author} {\bibinfo {author} {\bibfnamefont {M.}~\bibnamefont
  {Bernaschi}}, \bibinfo {author} {\bibfnamefont {S.}~\bibnamefont
  {Melchionna}}, \bibinfo {author} {\bibfnamefont {S.}~\bibnamefont {Succi}},
  \bibinfo {author} {\bibfnamefont {M.}~\bibnamefont {Fyta}}, \bibinfo {author}
  {\bibfnamefont {E.}~\bibnamefont {Kaxiras}}, \ and\ \bibinfo {author}
  {\bibfnamefont {J.~K.}\ \bibnamefont {Sircar}},\ }\href@noop {} {\bibfield
  {journal} {\bibinfo  {journal} {Computer Physics Communications}\ }\textbf
  {\bibinfo {volume} {180}},\ \bibinfo {pages} {1495} (\bibinfo {year}
  {2009})}\BibitemShut {NoStop}%
\bibitem [{\citenamefont {Feng}\ \emph {et~al.}(2019)\citenamefont {Feng},
  \citenamefont {Boivin}, \citenamefont {Jacob},\ and\ \citenamefont
  {Sagaut}}]{feng2019hybrid}%
  \BibitemOpen
  \bibfield  {author} {\bibinfo {author} {\bibfnamefont {Y.}~\bibnamefont
  {Feng}}, \bibinfo {author} {\bibfnamefont {P.}~\bibnamefont {Boivin}},
  \bibinfo {author} {\bibfnamefont {J.}~\bibnamefont {Jacob}}, \ and\ \bibinfo
  {author} {\bibfnamefont {P.}~\bibnamefont {Sagaut}},\ }\href@noop {}
  {\bibfield  {journal} {\bibinfo  {journal} {Journal of Computational
  Physics}\ }\textbf {\bibinfo {volume} {394}},\ \bibinfo {pages} {82}
  (\bibinfo {year} {2019})}\BibitemShut {NoStop}%
\bibitem [{\citenamefont {Montessori}\ \emph {et~al.}(2018)\citenamefont
  {Montessori}, \citenamefont {Lauricella}, \citenamefont {La~Rocca},
  \citenamefont {Succi}, \citenamefont {Stolovicki}, \citenamefont {Ziblat},\
  and\ \citenamefont {Weitz}}]{montessori2018regularized}%
  \BibitemOpen
  \bibfield  {author} {\bibinfo {author} {\bibfnamefont {A.}~\bibnamefont
  {Montessori}}, \bibinfo {author} {\bibfnamefont {M.}~\bibnamefont
  {Lauricella}}, \bibinfo {author} {\bibfnamefont {M.}~\bibnamefont
  {La~Rocca}}, \bibinfo {author} {\bibfnamefont {S.}~\bibnamefont {Succi}},
  \bibinfo {author} {\bibfnamefont {E.}~\bibnamefont {Stolovicki}}, \bibinfo
  {author} {\bibfnamefont {R.}~\bibnamefont {Ziblat}}, \ and\ \bibinfo {author}
  {\bibfnamefont {D.}~\bibnamefont {Weitz}},\ }\href@noop {} {\bibfield
  {journal} {\bibinfo  {journal} {Computers \& Fluids}\ }\textbf {\bibinfo
  {volume} {167}},\ \bibinfo {pages} {33} (\bibinfo {year} {2018})}\BibitemShut
  {NoStop}%
\bibitem [{\citenamefont {Montessori}\ \emph {et~al.}(2023)\citenamefont
  {Montessori}, \citenamefont {Lauricella}, \citenamefont {Tiribocchi},
  \citenamefont {Durve}, \citenamefont {La~Rocca}, \citenamefont {Amati},
  \citenamefont {Bonaccorso},\ and\ \citenamefont
  {Succi}}]{montessori2023thread}%
  \BibitemOpen
  \bibfield  {author} {\bibinfo {author} {\bibfnamefont {A.}~\bibnamefont
  {Montessori}}, \bibinfo {author} {\bibfnamefont {M.}~\bibnamefont
  {Lauricella}}, \bibinfo {author} {\bibfnamefont {A.}~\bibnamefont
  {Tiribocchi}}, \bibinfo {author} {\bibfnamefont {M.}~\bibnamefont {Durve}},
  \bibinfo {author} {\bibfnamefont {M.}~\bibnamefont {La~Rocca}}, \bibinfo
  {author} {\bibfnamefont {G.}~\bibnamefont {Amati}}, \bibinfo {author}
  {\bibfnamefont {F.}~\bibnamefont {Bonaccorso}}, \ and\ \bibinfo {author}
  {\bibfnamefont {S.}~\bibnamefont {Succi}},\ }\href@noop {} {\bibfield
  {journal} {\bibinfo  {journal} {Journal of Computational Science}\ }\textbf
  {\bibinfo {volume} {74}},\ \bibinfo {pages} {102165} (\bibinfo {year}
  {2023})}\BibitemShut {NoStop}%
\bibitem [{\citenamefont {Parsa}\ and\ \citenamefont
  {Wagner}(2020)}]{parsa2020large}%
  \BibitemOpen
  \bibfield  {author} {\bibinfo {author} {\bibfnamefont {M.~R.}\ \bibnamefont
  {Parsa}}\ and\ \bibinfo {author} {\bibfnamefont {A.~J.}\ \bibnamefont
  {Wagner}},\ }\href@noop {} {\bibfield  {journal} {\bibinfo  {journal}
  {Physical Review Letters}\ }\textbf {\bibinfo {volume} {124}},\ \bibinfo
  {pages} {234501} (\bibinfo {year} {2020})}\BibitemShut {NoStop}%
\bibitem [{\citenamefont {D’Orazio}\ and\ \citenamefont
  {Succi}(2003)}]{d2003boundary}%
  \BibitemOpen
  \bibfield  {author} {\bibinfo {author} {\bibfnamefont {A.}~\bibnamefont
  {D’Orazio}}\ and\ \bibinfo {author} {\bibfnamefont {S.}~\bibnamefont
  {Succi}},\ }in\ \href@noop {} {\emph {\bibinfo {booktitle} {International
  Conference on Computational Science}}}\ (\bibinfo {organization} {Springer},\
  \bibinfo {year} {2003})\ pp.\ \bibinfo {pages} {977--986}\BibitemShut
  {NoStop}%
\bibitem [{\citenamefont {Guo}, \citenamefont {Shi},\ and\ \citenamefont
  {Zheng}(2002)}]{guo2002coupled}%
  \BibitemOpen
  \bibfield  {author} {\bibinfo {author} {\bibfnamefont {Z.}~\bibnamefont
  {Guo}}, \bibinfo {author} {\bibfnamefont {B.}~\bibnamefont {Shi}}, \ and\
  \bibinfo {author} {\bibfnamefont {C.}~\bibnamefont {Zheng}},\ }\href@noop {}
  {\bibfield  {journal} {\bibinfo  {journal} {International Journal for
  Numerical Methods in Fluids}\ }\textbf {\bibinfo {volume} {39}},\ \bibinfo
  {pages} {325} (\bibinfo {year} {2002})}\BibitemShut {NoStop}%
\end{thebibliography}

%

\end{document}